\documentclass[sigconf]{acmart}

\usepackage{makecell}
\usepackage{bm}
\usepackage{enumitem}
\setlist{nolistsep}
\usepackage{multirow}
\usepackage{float}
\usepackage{tabularx}
\usepackage{pifont}
\usepackage{booktabs}
\usepackage{caption}
\usepackage{colortbl}
\usepackage[subrefformat=parens,skip=0pt]{subcaption}
\usepackage{marginnote}
\usepackage[detect-all]{siunitx}
\usepackage{xspace}
\usepackage[bb=boondox,bbscaled=.95,cal=boondoxo]{mathalfa} 

\newcommand\ie{i.\,e.\xspace}
\newcommand\eg{e.\,g.\xspace}

\newcommand\US{U.S.\xspace}

\newcommand{\X}{$\mathbb{X}$\xspace}
\newcommand{\var}[1]{\mathit{#1}}

\def\sym#1{\ifmmode^{#1}\else\(^{#1}\)\fi}

\AtBeginDocument{%
  }

\copyrightyear{2026}
\acmYear{2026}
\setcopyright{cc}
\setcctype{by}
\acmConference[CHI '26]{Proceedings of the 2026 CHI Conference on Human Factors in Computing Systems}{April 13--17, 2026}{Barcelona, Spain}
\acmBooktitle{Proceedings of the 2026 CHI Conference on Human Factors in Computing Systems (CHI '26), April 13--17, 2026, Barcelona, Spain}
\acmPrice{}
\acmDOI{10.1145/3772318.3790524}
\acmISBN{979-8-4007-2278-3/2026/04}

\begin{document}

\title{Request a Note: How the Request Function Shapes X's Community Notes System}


\author{Yuwei Chuai}
\orcid{0000-0001-6181-7311}
\affiliation{
  \institution{University of Luxembourg}
  \city{Luxembourg}
  \country{Luxembourg}}
\email{yuwei.chuai@uni.lu}

\author{Shuning Zhang}
\orcid{0000-0002-4145-117X}
\affiliation{
  \institution{Tsinghua University}
  \city{Beijing}
  \country{China}}
\email{zsn23@mails.tsinghua.edu.cn}

\author{Ziming Wang}
\orcid{0000-0003-0564-8757}
\affiliation{%
  \institution{University of Luxembourg}
  \city{Esch-sur-Alzette}
  \country{Luxembourg}}
\affiliation{%
  \institution{Chalmers University of Technology}
  \city{Gothenburg}
  \country{Sweden}}
\email{ziming.wang.001@student.uni.lu}

\author{Xin Yi}
\orcid{0000-0001-8041-7962}
\affiliation{
  \institution{Tsinghua University}
  \city{Beijing}
  \country{China}}
\email{yixin@tsinghua.edu.cn}

\author{Mohsen Mosleh}
\orcid{0000-0001-7313-5035}
\affiliation{
  \institution{University of Oxford}
  \city{Oxford}
  \country{United Kingdom}}
\email{mohsen.mosleh@oii.ox.ac.uk}

\author{Gabriele Lenzini}
\orcid{0000-0001-8229-3270}
\affiliation{
  \institution{University of Luxembourg}
  \city{Luxembourg}
  \country{Luxembourg}}
\email{gabriele.lenzini@uni.lu}

\begin{abstract}
\X's Community Notes is a crowdsourced fact-checking system. To improve its scalability, \X introduced ``Request Community Note'' feature, enabling users to solicit fact-checks from contributors on specific posts. Yet, its implications for the system---what gets checked, by whom, and with what quality---remain unclear. Using \num{98685} requested posts and their associated notes, we evaluate how requests shape the Community Notes system. We find that requested posts with higher GPT-estimated misleadingness and from authors with greater misinformation exposure are more likely to receive notes. Conversely, requested political posts (vs. non-political) are less likely to receive notes. We also observe partisan asymmetries: posts from Republicans are more likely to receive notes than those from Democrats. Although only 12\% of requested posts receive request-fostered notes from top contributors, these notes are rated as more helpful and less polarized than others, partly reflecting top contributors' selective fact-checking of misleading posts. Our findings highlight both the limitations and promise of requests for scaling high-quality community-based fact-checking.
\end{abstract}


\begin{teaserfigure}
  \centering
  \includegraphics[width=.8\textwidth]{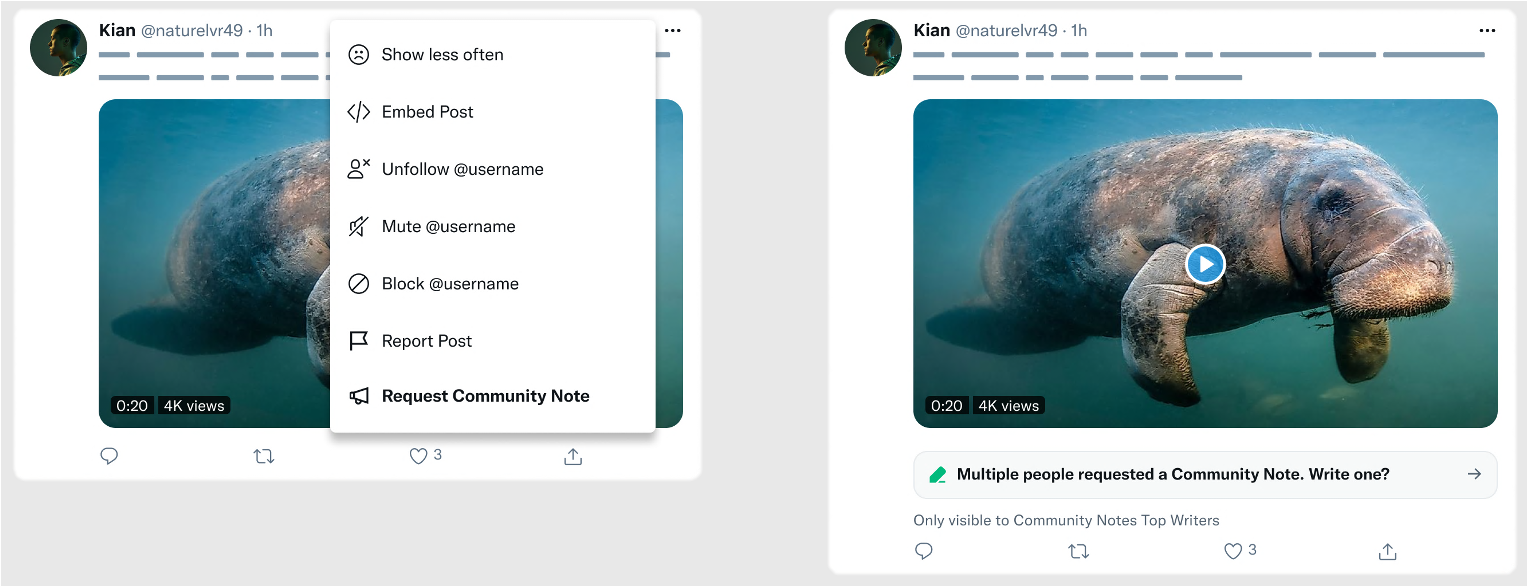}
  \caption{The introduction of ``Request Community Note'' feature on \X, sourced from the official Community Notes website~\cite{cn2025request}. It allows users to request a note on a post they believe would benefit from additional context (see the left part of the figure). When a sufficient number of requests are submitted, a subset of Community Notes contributors (i.e., top writers) will see an alert below the corresponding requested post, and can choose to propose a note (see the right part of the figure).}
  \label{fig:cn_request}
  \Description{The introduction of "Request Community Note" feature on X, sourced from the official Community Notes website. It allows users to request a note on a post they believe would benefit from additional context (see the left part of the figure). When a sufficient number of requests are submitted, a subset of Community Notes contributors (i.e., top writers) will see an alert below the corresponding requested post, and can choose to propose a note (see the right part of the figure).}
\end{teaserfigure}

\maketitle
\section{Introduction}
Designing effective countermeasures to identify falsehoods and curb the spread of misleading posts (\ie, posts carrying misinformation) on social media platforms is still a challenge. Years ago, the social media platform \X (formerly Twitter) adopted \emph{Community Notes}, a crowdsourced fact-checking system designed to address misleading posts and provide users with helpful additional context.\footnote{Here, misinformation refers to false or misleading information that is contradicted by empirical evidence or inconsistent with common or expert consensus, without imposing assumptions on the intent or the format of the content~\cite{chuai2024roll}. It serves as an umbrella term including not only factually incorrect content but also disinformation and fake news---both deliberately fabricated to mislead others and/or cause harms~\cite{pennycook2021psychology,chuai2024roll,ecker2022psychological}. Prior work further shows that factually accurate yet misleading content can be more influential than outright falsehoods~\cite{allen2024quantifying}. In the Community Notes system, the criteria used for labeling misleading posts align closely with our definition of misinformation (see details in Suppl.~\ref{sec:misleading_criteria}).} This initiative marks the first large-scale implementation of community-based fact-checking by a major social media platform~\cite{chuai2024roll,augenstein2025community}. A subset of users---called \emph{contributors}---are periodically admitted into the program if they meet specific eligibility requirements, such as having no recent violations of \X's rules. Contributors begin by rating the helpfulness of existing notes, and only after demonstrating reliability can they unlock the ability to write notes themselves. All notes are evaluated by a note selection algorithm, and then those deemed helpful by the algorithm are displayed directly on the relevant posts~\cite{cn2025signing}. Inspired by this program, other social media providers, including YouTube, Meta, and TikTok, are beginning to explore similar community-based moderation models~\cite{tiktok2025testing,youtube2024testing,meta2025testing}. For example, Meta is piloting the Community Notes program across Facebook, Instagram, and Threads in the \US, while TikTok plans to offer a similar feature called Footnotes to provide context and corrections for misleading posts. Given the rapid expansion of community-based fact-checking system, it is crucial to understand how the system functions in the real world.

Recent work has shown that \X's Community Notes program can reliably attach helpful notes to the corresponding misleading posts~\cite{wojcik2022birdwatch,allen2024characteristics}. Moreover, community fact-checks are perceived as more trustworthy than expert fact-checks, can effectively reduce the spread of misleading posts, and even prompt authors to delete their problematic content~\cite{drolsbach2024community,chuai2024community.new,wojcik2022birdwatch,chuai2025community}. However, the Community Notes system, in its current implementation on \X, still faces limitations, such as the small number of displayed notes relative to all generated notes and restrictions on who can author them~\cite{chuai2024community.new}. 

To foster note writing and create a more interactive and responsible community, \X recently implemented a key innovation in Community Notes system---the \emph{Request Community Note} feature (as illustrated in Fig.~\ref{fig:cn_request}), allowing broader registered users to signal posts they believe would benefit from additional context. This mechanism enables non-contributor users to participate in the moderation process while guiding contributors toward posts where notes are likely to be needed. This request feature is expected to increase the involvement and the interest of users in checking facts, which is crucial in the current media landscape, where misinformation can spread rapidly. However, it remains unclear whether this mechanism effectively improves the Community Notes system by helping more misleading posts receive notes and enhancing the quality of the resulting notes.

\vspace{1mm}
\noindent
\textbf{Research questions.}
In this study, we examine the newly introduced Request Community Note feature on \X, tracing the process from the submission of requests to the evaluation of the associated notes. The workflow involves requestors flagging posts (\ie, \emph{requested posts}) for note writers, note writers selecting requested posts to propose community notes, and the note selection algorithm evaluating the generated notes. Notably, note writers retain autonomy: they can independently choose any post to fact-check and propose community notes, regardless of whether they see request alerts or not. Moreover, the note selection algorithm evaluates community notes along two key dimensions: \emph{helpfulness}, indicating the consensus of perceived helpfulness among heterogeneous contributors, and \emph{polarization}, capturing the degree of variation in ratings. Accordingly, we organize our investigation around three research questions (RQs) to explore how note writers select requested posts to fact-check, the extent to which their selections are influenced by requests, and how community notes fostered by user requests (\ie, \emph{request-fostered notes}) differ from other notes in helpfulness and polarization according to the note selection algorithm:
\vspace{1mm}
\begin{itemize}[leftmargin=*]
    \item \textbf{RQ1:} Among requested posts, which ones actually receive community notes, and how do their content and author characteristics shape this likelihood?
    \item \textbf{RQ2:} To what extent do user requests foster the creation of community notes, and how do they influence contributors' selection of posts?
    \item \textbf{RQ3:} Are community notes fostered by user requests more or less helpful and polarized than other notes, and what factors explain these differences?
\end{itemize}
\vspace{1mm}
\textbf{Data and methods.}
To address our research questions, we collect all available Community Notes data and associated posts that had received at least five requests as of May 20, 2025. We then specify a logistic regression model to estimate the likelihood of a post receiving a community note based on a wide range of characteristics extracted from posts, such as topics and estimated misleadingness, and their authors, such as verified status and misinformation exposure scores~\cite{mosleh2022measuring} (RQ1). In addition, we analyze the extent to which requests foster the creation of community notes and how requests can influence the choices of note writers (RQ2). Finally, we replicate the note selection algorithm to evaluate request-fostered notes compared to other notes and explore factors affecting the evaluation results (RQ3).

\vspace{1mm}
\noindent
\textbf{Contributions.} 
Overall, our study results in three main findings. (i) The likelihood that a requested post receives a community note is associated with the post's content and its author. For instance, posts related to entertainment or finance, containing claim-oriented content, with higher estimated misleadingness, or from accounts with higher misinformation exposure scores, are more likely to receive community notes. Our findings suggest that Community Notes contributors and requestors may have different selection patterns. (ii) With more than half (53.6\%) of requested posts receiving community notes, only an estimated 12.1\% of those are likely fostered by user requests, showing that the effect of requesting community notes in fostering note generation is, so far, limited. Nonetheless, requests tend to shift contributors' attention toward political content, while prioritizing posts with higher misleadingness (as measured by GPT-4.1 from OpenAI). (iii) Request-fostered notes generated by top writers are evaluated as more helpful and less polarized than other notes, a pattern that likely reflects writers' selective prioritization of requested posts with higher levels of misleadingness. To our knowledge, our study provides the first empirical evaluation of this novel request mechanism, offering insights into both the opportunities and challenges of scaling crowdsourced fact-checking systems.

\section{Background and Related Work}

The spread of misleading posts on social media platforms---including \X (formerly Twitter), TikTok, and Facebook---remains a significant societal problem, with far-reaching and often harmful consequences~\cite{lazer2018science,ecker2024misinformation}. For example, the widespread misinformation during political elections is particularly alarming, as exposure to and belief in false narratives during elections can distort public opinion, erode trust in democratic institutions, and destabilize societies~\cite{ecker2024misinformation,aral2019protecting,vosoughi2018spread,green2022online}. Beyond electoral contexts, misinformation can exacerbate social divisions and intensify polarization in response to global pandemics and claim change~\cite{ecker2022psychological,ecker2024misinformation}. Given these multifaceted risks, social media providers face increasing pressure to adopt transparent content moderation policies and implement effective and scalable interventions to identify misleading posts, correct falsehoods, and ensure users have access to reliable information~\cite{donovan2020social,lazer2018science,drolsbach2024content,bar2023new}.

\subsection{Misinformation Identification and Fact-Checking}

To identify misinformation, traditional human-driven fact-checking often relies on professional experts and third-party organizations. However, the scarcity of expert fact-checkers, coupled with the relentless pace of online content, hampers the timely and scalable implementation of professional fact-checking~\cite{chuai2024roll,augenstein2025community}. Furthermore, despite its depth and rigor, professional fact-checking has faced criticism, as experts are sometimes perceived as biased in selecting which claims to review---raising concerns about agenda-setting and potentially eroding public trust in their assessments~\cite{chuai2025political,lee2023fact}. To address the limitations of professional fact-checking, crowdsourced (or community-based) fact-checking approaches have been proposed, leveraging the wisdom of crowds to achieve broader and faster coverage of online content~\cite{allen2021scaling,godel2021moderating}. Early explorations of crowdsourced verification demonstrate that journalists and online volunteer communities can collaboratively assess information accuracy and provide real-time event coverage during emergencies~\cite{dailey2014journalists}. Furthermore, hybrid crowdsourcing processes that involve both local and remote participants help enhance the completeness of information~\cite{agapie2015crowdsourcing}. Although individual assessments may be subject to bias and noise~\cite{allen2022birds}, aggregated judgments---even from relatively small crowds---have been shown to be reliable and comparable in accuracy to expert evaluations~\cite{bhuiyan2020investigating,epstein2020will,pennycook2019fighting,saeed2022crowdsourced,allen2021scaling}. Moreover, crowdsourced assessments have the potential to mitigate public trust issues associated with expert fact-checks~\cite{zhang2024profiling,drolsbach2024community,chuai2025political}. Together, these studies provide important foundations for the development of community-based fact-checking systems.

Meanwhile, automatic fact-checking powered by machine learning models continues to develop, offering increasingly sophisticated tools for identifying misinformation. Traditional detection models are typically trained on surface-level content or contextual features such as linguistic patterns, semantic similarity, propagation structures, or user metadata~\cite{chen2024combating,chuai2024roll}. Although these approaches have proven useful in identifying repeated misinformation or content with distinctive stylistic markers, particularly in constrained domains or benchmark datasets, they often struggle with more subtle, novel, or context-dependent cases. In particular, such models lack the ability to reason over evidence, integrate external knowledge sources, or adapt quickly to evolving narratives. As a result, their performance in real-world fact-checking scenarios often falls short, especially when confronted with emerging claims that deviate from the data on which they were trained~\cite{glockner2022missing}.

Recent advances in Large Language Models (LLMs), such as OpenAI's GPT series, introduce novel opportunities to overcome the shortcomings of traditional machine learning models in fact-checking. LLMs exhibit strong contextual understanding and can simulate human-like reasoning processes, allowing them to evaluate claims not only on textual cues but also in relation to broader world knowledge. They can flexibly incorporate external evidence through retrieval-augmented generation, adapt to diverse domains with minimal task-specific training, and generate explanations that improve transparency in the fact-checking process~\cite{chen2024combating,warren2025show}. Previous work suggests that LLMs can achieve human-comparable performance in misinformation detection tasks~\cite{choi2024fact,deverna2024fact,costabile2025assessing,rathje2024gpt,majer2024claim}. For instance, ChatGPT has been shown to achieve an accuracy of approximately 90\% in identifying false headlines, demonstrating the potential of LLMs to support large-scale misinformation detection~\cite{deverna2024fact}. At the same time, LLM-based fact-checking is not without challenges. LLMs tend to adopt a more structured evaluation strategy, whereas human annotators often display greater variability in their use of evaluative criteria, particularly for borderline or ambiguous claims~\cite{costabile2025assessing}. More critically, their influence on user perception can be harmful: when LLMs mislabel true claims as false, they risk reducing belief in accurate information; conversely, when they express uncertainty about false claims, they may inadvertently increase belief in misinformation~\cite{deverna2024fact,lu2025understanding}. These limitations underscore that, while LLMs can perform well in detecting misinformation, human judgment remains essential for deciding what content to surface to users, helping to mitigate unintended harms.

\subsection{Crowdsourced Fact-Checking in Practice: X's Community Notes}

Given the necessity of human-centered approaches and the demonstrated promise of crowdsourced fact-checking, \X introduced Community Notes, a system that engages diverse users in collaboratively annotating potentially misleading content at scale~\cite{cn2025signing}. Specifically, Community Notes contributors can append contextual information or corrections to potentially misleading posts, offering alternative perspectives or clarifications. Contributors are required to cite external sources in their community notes. These notes are continuously evaluated through peer ratings from other contributors. However, featuring community notes based on majority-based aggregation rules makes the system vulnerable to manipulation or one-side ratings from ideologically homogeneous raters. Research shows that politically balanced groups produce more reliable and accurate evaluations than homogeneous ones~\cite{allen2021scaling,godel2021moderating}. Therefore, \X develops a bridging algorithm to feature objectively informative community notes that are subjectively perceived to be helpful across heterogeneous contributors~\cite{wojcik2022birdwatch}. Beyond contributors, prior research has also explored on-demand fact-checking models that enable ordinary users to request fact-checks for content they wish to have checked~\cite{kriplean2014integrating}. In line with this idea, all users on \X can now participate indirectly through the recently launched Request Community Note function, which allows any account with a verified phone number to submit requests for posts they believe would benefit from a community note. When enough requests are submitted, \emph{top writers} can see alerts highlighting these posts and can choose to write a note in response (see the illustrations in Fig.~\ref{fig:cn_request}). This division of roles---between note writers, note raters, and requestors---illustrates how Community Notes distributes responsibilities across different layers of the user base, thereby expanding participation beyond the core contributor group.

Research has shown that Community Notes system can successfully identify misleading content, reduce its spread, and even pressure authors to delete problematic posts~\cite{wojcik2022birdwatch,martel2024fact,allen2024characteristics,chuai2024community.new}. Additionally, community-based annotations are often perceived as more trustworthy than professional fact-checks~\cite{drolsbach2024community}. These encouraging findings suggest that Community Notes is promising in addressing misinformation on social media platforms and have sparked growing scholarly interest in understanding the dynamics of the system and identifying opportunities to refine and enhance its design~\cite{augenstein2025community}. For instance, despite the relatively large volume of notes generated by contributors, only a small fraction---around 10\%---achieve sufficient support from diverse users and are displayed to users~\cite{de2025supernotes,chuai2024community.new}. At the same time, the display process is often not fast enough to intervene before misleading posts have already begun to spread widely~\cite{chuai2024community.new,chuai2024roll}. To address these limitations, recent work has explored two complementary directions. 

First, researchers and practitioners have turned to technological augmentation, using Large Language Models (LLMs) to help scale human judgment by assisting contributors in drafting and refining notes~\cite{li2025scaling}. For example, the Supernotes approach generates AI-synthesized summaries that integrate insights from multiple existing contributor notes with the goal of fostering consensus among diverse raters~\cite{de2025supernotes}. Second, \X has expanded user participation by introducing the request function, which allows any user with a verified account to flag posts that they believe should receive a community note. This structural expansion extends the influence of the wider user base beyond the core contributor group, potentially increasing the coverage and responsiveness of the system. However, despite its significance, this new function has not yet been systematically studied, leaving open questions about its effects on content selection for fact-checking, note generation, and the quality of resulting notes. Addressing these questions represents the primary goal of this study.

\section{Data and Methods}
In this section, we first describe our data collection process, including the posts, requests, and associated community notes. We then outline the post content and author characteristics extracted for analysis, including GPT–based evaluations of post misleadingness as a proxy for misinformation risk. Next, we examine external source domains cited in community notes, assessing them in terms of information quality and political bias. Since request alerts are only visible to top writers, we also identify top note writers following the criteria provided on the official Community Notes website. Finally, we replicate the note selection algorithm to evaluate the helpfulness and polarization of notes, enabling us to compare request-fostered notes with those generated independently of requests.

\subsection{Data Collection}
To address our research questions, we downloaded the complete set of available Community Notes data---including requests, notes, ratings, and note status histories---on May 20, 2025, when \X began releasing request data to the public~\cite{cn2025downloading}. Specifically, we collected \num{5888351} requests submitted by \num{1689152} unique users since the launch of the request feature on July 18, 2024. In addition, the dataset includes \num{1787609} notes that have received \num{149646500} ratings since the introduction of Community Notes program.

Subsequently, we used the full-archive search endpoint of the \X Pro API to collect the requested posts. In total, the submitted requests target \num{2427451} unique posts. However, according to the platform guidelines at the time of data collection, a post must receive at least 5 requests to be surfaced to top writers, indicating a threshold for which posts are eligible for community evaluation~\cite{cn2025request}. In our dataset, \num{154090} posts meet this criterion, representing only 6.3\% of all requested posts. In contrast, the requests submitted to these posts account for more than half (52.6\%) of the total requests. This suggests that, while users submitted a large volume of requests, attention was concentrated on a small set of posts. Based on the corresponding IDs of these eligible posts, we successfully retrieved \num{133351} posts, of which \num{98685} (74\%) are in English and were authored by \num{22915} unique accounts. We restrict our analysis to the collected posts in English and focus on the request feature in the \US context. Next, we introduce how to extract the characteristics of posts and their authors for our subsequent analysis (see summary statistics in Table~\ref{tab:data_summary}).

\begin{table*}
\centering
\caption{Overview of requested posts. The three columns are for all requested posts, posts with community notes, and posts without community notes, respectively. The mean values for continuous variables (standard deviations in parentheses), percentages for dummy variables, or count numbers for count variables are reported.}
\begin{tabular}{l*{3}{c}}
\toprule
&{(1)} & {(2)}& {(3)}\\
&{All}&{With notes}&{Without notes}\\
\midrule
\# Requested posts&{98,685}&{52,862 (53.6\%)}&{45,823 (46.4\%)}\\
\# Post authors&{22,915}&{15,575}&{12,779}\\
\underline{Post content characteristics}\\
\quad Sentiments: Positive& {0.131 (0.244)} & {0.134 (0.246)} & {0.127 (0.241)}\\
\quad Sentiments: Negative& {0.460 (0.335)} & {0.448 (0.335)} & {0.474 (0.335)}\\
\quad Topics: Politics& {37.0\%} & {32.4\%} & {42.4\%}\\
\quad Topics: Science \& Technology& {13.5\%} & {12.8\%} & {14.2\%}\\
\quad Topics: Entertainment& {26.9\%} & {28.0\%} & {25.6\%}\\
\quad Topics: Finance \& Business& {32.6\%} & {33.9\%} & {31.2\%}\\
\quad Content types: Claim& {0.687 (0.318)} & {0.687 (0.324)} & {0.687 (0.312)}\\
\quad Content types: Opinion& {0.489 (0.346)} & {0.475 (0.347)} & {0.506 (0.344)}\\
\quad GPT misleadingness& {0.398 (0.318)} & {0.405 (0.324)} & {0.389 (0.310)}\\
\quad Media& {65.4\%} & {68.4\%} & {62.0\%}\\
\underline{Post author characteristics}\\
\quad Account types: Blue& {72.5\%} & {72.3\%} & {72.6\%}\\
\quad Account types: Business& {6.6\%} & {6.4\%} & {6.7\%}\\
\quad Account types: Government& {7.2\%} & {6.5\%} & {7.9\%}\\
\quad Account age& {3,156.145 (1,997.934)} & {3,078.018 (1,998.710)} & {3,246.273 (1,993.259)}\\
\quad Followers& {6,828,060} & {6,561,490} & {7,135,578}\\
\quad Followees& {13,347} & {11,081} & {15,960}\\
\quad Misinformation exposure score& {0.546 (0.139)} & {0.555 (0.143)} & {0.536 (0.135)}\\
\quad Partisan score& {\num{-0.035} (0.767)} & {0.023 (0.768)} & {\num{-0.095} (0.761)}\\
\bottomrule
\end{tabular}
\label{tab:data_summary}
\Description{A table for the overview of post content and the authors' characteristics. There are three columns for all requested posts, posts with community notes, and posts without community notes, respectively. The rows of the table include the number of posts, the number of authors, statistics of post content characteristics, and statistics of post author characteristics. Each row has three values corresponding to the three columns.}
\end{table*}

\subsection{Characteristics of Post Content}
\subsubsection{Sentiments}
Sentiments are an important factor that influences the dissemination of online content~\cite{chuai2022anger,robertson2023negativity}. We extract sentiments in the posts content using a state-of-the-art Twitter-roBERTa-base sentiment classification model (2022 updated), which is available on HuggingFace.\footnote{\url{https://huggingface.co/cardiffnlp/twitter-roberta-base-sentiment-latest}} This model was trained on 124 million posts created from January 2018 to December 2021, and fine-tuned for sentiment analysis. It achieves superior predictive performance on the TweetEval benchmark, with a reported macro-recall of 0.737 on the sentiment classification task~\cite{loureiro2022timelms}. The model outputs probability estimates for three sentiment categories: positive, neutral, and negative, with the sum of the probabilities across these categories equaling to 1. On average, requested posts exhibit strongest negative sentiment (mean of \num{0.460}), followed by neutral sentiment (mean of \num{0.409}; $t(\var{negative}, \var{neutral})=$ \num{28.651}, $p<$ \num{0.001}) and positive sentiment (mean of \num{0.131}; $t(\var{negative}, \var{positive})=$ \num{199.208}, $p<$ \num{0.001}). In this study, we focus on affectively charged content; thus,  to avoid multicollinearity in our empirical regression analyses, we retain the probabilities of positive and negative sentiments for each collected post and omit the neutral probability. 

\subsubsection{Topics}
Online misinformation often leverages sentiments to enhance its virality and tends to concentrate on specific topics---most notably, \eg, politics~\cite{vosoughi2018spread,chuai2022anger}. \X has provided domain labels for each post in the collected dataset. Following prior work~\cite{chuai2024roll}, we extract these domains from the requested posts and group them into four broad topics: Politics, Science \& Technology, Entertainment, and Finance \& Business. Notably, a single post may be associated with multiple topics. Politics (37\%) is the most frequent topic in the requested posts, followed by Finance \& Business (32.6\%), Entertainment (26.9\%), and Science \& Technology (13.5\%).

\subsubsection{Content types and misleadingness}
Traditional expert-based fact-checking and emerging LLM-based misinformation detection methods primarily aim to identify checkworthy, objective, and verifiable claims, subsequently evaluating their truthfulness~\cite{chuai2025political,costabile2025assessing}. These approaches focus on factual claims and often overlook opinion-based content that may contain misleading interpretations~\cite{ni2024afacta}. In contrast, the Community Notes system relies on crowd consensus to assess the misleadingness of selected posts and surface community fact-checks, with the potential to address both factual inaccuracies and misleading opinions (see the misleading criteria in Suppl.~\ref{sec:misleading_criteria}). Given this, we distinguish between \emph{factual claims}, which explicitly presents some verifiable facts, and \emph{personal opinions}, which are more subjective content to express users' personal judgments (see full definitions in Supp.~\ref{sec:gpt_prompts}). This distinction is important for understanding how requests and community notes select verifiable content and opinionated posts. Notably, recent critiques also highlight that a substantial portion of community notes have been attached to opinion-based posts and published for debate rather than moderation~\cite{razuvayevskaya2025timeliness}. To assess whether requestors and contributors target content that is plausibly misleading, rather than merely content they disagree with, we analyze the misleadingness of requested posts. 

Given that recent work demonstrates the comparable performance of LLMs, such as GPT, to human evaluations in text quality and their potential applications in large-scale empirical analysis~\cite{chiang2023large,rathje2024gpt,deverna2024fact}, we use GPT-4.1, a state-of-the-art LLM model at the time of this study, to classify content types (claims vs. opinions) and to estimate the misleadingness of requested posts. To ensure that the model's outputs align with the scope of our analysis, we provide GPT a task-specific prompt incorporating the definitions and criteria for misleading information, factual claims, and opinions (see full prompt in Suppl.~\ref{sec:gpt_prompts}). The GPT-estimated scores for claim, opinion, and misleadingness for each post range from 0 to 1, with higher values indicating a stronger presence of the corresponding attribute. Additionally, a post can contain both claims and opinions, and their estimated scores are not required to sum to 1.

To assess the reliability of GPT-generated annotations, we first randomly select 200 posts from the requested dataset and conduct a manual validation using a small crowd. Specifically, seventeen university-level student assistants are recruited and trained on the relevant definitions and criteria to annotate whether each assigned posts contain claims, opinions, and misleading information. Each post is independently reviewed by a minimum of six assistants to reduce individual-level noise and ensure robust aggregation, with final annotations determined by averaging their ratings. For comparison, we transform the GPT-estimated scores into binary classes using a threshold of 0.5, and compare the GPT-generated labels with the assistant-annotated labels. We find that the GPT-based approach demonstrates strong performances for claims ($\var{Accuracy}=$ 0.840, $\var{Weighted~F1}=$ 0.838), opinions ($\var{Accuracy}=$ 0.805, $\var{Weighted~F1}=$ 0.810), and misleadingness ($\var{Accuracy}=$ 0.800, $\var{Weighted~F1}=$ 0.800). These results suggest that the GPT-generated outputs are comparable to the aggregated human judgments, supporting their use in our large-scale analysis (see its validation with expert fact-checks from PolitiFact in Suppl.~\ref{sec:gpt_politifact}). Overall, the requested posts are more likely to contain claims (mean of 0.687), compared to opinions (mean of 0.489; $t=$ \num{111.269}, $p<$ \num{0.001}). Additionally, the GPT-estimated misleadingness is, on average, \num{0.398} in the requested posts.\footnote{We use GPT to annotate the 200 sampled posts twice and assess the consistency of its outputs across the two rounds. The outputs show strong agreement across all three variables: claims (Cohen's $\kappa=0.920$), opinions (Cohen's $\kappa=0.828$), and misleadingness (Cohen's $\kappa=0.816$). In addition, we further validate the reliability of GPT-based approach using expert fact-checks from PolitiFact ($\var{Accuracy}=$ 0.801, $\var{Weighted~F1}=$ 0.857; see details in Suppl.~\ref{sec:gpt_politifact}).}

\subsubsection{Media posts}
Media elements, such as images or videos, can increase user engagement with associated posts, making media-based misinformation potentially more viral and harmful than text-only content~\cite{chuai2024roll,pataranutaporn2025synthetic}. Additionally, the Community Notes team on \X implemented a media note feature, which allows helpful media notes to be displayed across all posts that contain the same media~\cite{cn2025media}. This feature can help curb the spread of misleading posts at the very beginning if they contain the same misleading media annotated by existing community notes, thereby increasing the responsiveness of the Community Notes system. By analyzing the media keys returned by the \X Pro API, we find that more than half of all requested posts contain media (65.4\%).

\subsection{Characteristics of Post Authors}
\subsubsection{Built-in characteristics of post authors}
We analyze several built-in features of the post authors obtained from the \X Pro API. The dataset includes in total \num{22915} unique post accounts who authored the requested posts. Each author account is categorized as either verified or unverified. Verified accounts are categorized as \emph{Blue}, \emph{Government}, or \emph{Business}: Blue accounts correspond to individual \X Premium subscribers, while government and business accounts are officially verified by \X. We find that the requested posts are predominantly authored by blue accounts (72.5\%), followed by unverified (13.8\%), government (7.2\%), and business (6.6\%) accounts. In addition, we consider the number of followers (mean of \num{6828060}; median of \num{326732}), the number of followees (mean of \num{1334676}; median of \num{1180}), and the account age, defined as the number of days from the account creation to the creation of the corresponding post. The average of the account ages in the requested posts is \num{3156} days, \ie, $\sim$ 9 years. These features provide insights into the authors' influence, reach, and longevity on the platform.

\subsubsection{Misinformation exposure score and partisan score}
Previous work developed a method to calculate users' exposure to misinformation from political elites on \X~\cite{mosleh2022measuring}. They assigned each elite (public figures and organizations) a ``falsity score'' based on the veracity of their statements, as determined by professional fact-checkers like PolitiFact. Users' misinformation exposure scores were then computed by averaging the falsity scores of the elites they followed on \X. This approach allows researchers to assess the relationship between users' exposure to elite misinformation and their sharing behaviors, as well as their estimated political ideologies. We estimate the political partisanship and misinformation exposure of the relevant post authors on \X using the API service provided by this study. Partisanship scores range from \num{-1} (Democrat) to \num{+1} (Republican) and are based on the number of Democratic and Republican public figures followed by each user. The misinformation exposure score, ranging from 0 to 1, reflects the proportion of followed public figures rated as false by PolitiFact. Out of \num{22915} post authors, we successfully collect misinformation exposure scores and partisan scores for \num{10492} (45.8\%) authors. These scores, together with other post and author characteristics, provide key context for analyzing which posts are more likely to attract community notes and how authors' influence and ideological orientations affect note generation. Summary statistics for the characteristics related to posts and their authors are reported in Table~\ref{tab:data_summary}.

\subsection{External Domain References in Community Notes}
We extract the external domains cited in the community notes based on string matching. To assess the quality and political bias of external domains referenced in the community notes, we rely on domain ratings compiled by~\citet{lin2023high}. This study aggregated six sets of expert news domain ratings using an ensemble approach that combined imputation and principal component analysis, resulting in a comprehensive set of quality scores for \num{11520} domains. The aggregated ratings provide a reliable measure of news quality, allowing us to evaluate the credibility of the information cited in community notes. In addition, we assess the political bias of external domains using Media Bias/Fact Check, a widely used online resource covering over \num{10000} media sources, journalists, politicians, and countries~\cite{solovev2025references}. Analogous to previous research~\cite{solovev2025references,kuuse2025crowdsourced}, we map bias categories to scores ranging from \num{-1} to 1. Specifically, the ``Least Biased'' and ``Pro-Science'' are coded as 0. The ``Left'' and ``Right'' are coded as \num{-1} and 1, respectively. The ``Left-Center'' and ``Right-Center'' are coded as \num{-0.5} and 0.5, respectively. Finally, we successfully assign domain quality scores and domain bias scores to external domains in \num{367378} community notes, accounting for 20.6\% of the total notes. Together, these measures allow us to evaluate both the reliability and ideological leaning of the information cited in the community notes.

\subsection{Identification of Top Writers}
Since only top note writers can view request alerts once the associated posts reach the request threshold~\cite{cn2025request}, it is essential to identify these top writers in order to evaluate whether requests actually drive note writing and coverage. The dataset, however, does not directly indicate top-writer status. To address this, we approximate the top writers using the criteria provided on the official Community Notes website: contributors must have at least 4\% of their notes rated ``Helpful''~\cite{cn2025top}. Based on this criterion, we identify \num{56113} top note writers, representing 22.2\% of all note writers (\num{433936}). Notably, the recognition of top-writer status also requires a writing impact score of 10 or higher, but this metric is not publicly available. Despite this limitation, our approach offers a conservative estimate that avoids underestimating the potential effect of requests on note generation.

\subsection{Replication of Note Selection Algorithm}
The Community Notes program employs a matrix-factorization-based approach to identify annotations that resonate across heterogeneous user groups~\cite{wojcik2022birdwatch}. We replicate this note selection algorithm to evaluate both the helpfulness and polarization of community notes~\cite{cn2025opensource}. The algorithm produces two key estimates: the \emph{note intercept}, which captures overall helpfulness indicating how broadly a note is judged to be useful across raters; and the \emph{note factor}, which, by contrast, measures polarization reflecting the extent to which contributor ratings diverge. Together, these metrics enable us to assess not only the quality of community-generated fact-checks but also the degree of consensus surrounding them. To validate our replication, we compare the note statuses generated by the reproduced algorithm with the actual production statuses. Each community note can receive one of the three statuses: Currently Rated Helpful, Needs More Ratings, and Currently Rated Not Helpful. Overall, 99.7\% of replicated statuses match their production counterparts, demonstrating the reliability of our reproduction.

\subsection{Estimation Models}
In this study, we employ a series of regression models and conduct empirical analyses using our observational dataset and pre-defined variables to address the research questions. 

For RQ1 and RQ2, we specify a logistic regression model to examine how contributors select posts for fact-checking:
\begin{equation}
\begin{aligned}
    logit(P(\var{Noted_{i}}))=\, & \beta_{0} + \bm{\beta_{1}'\var{X_{i}^{Post}}} + \bm{\beta_{2}'\var{X_{i}^{Poster}}},
\end{aligned}
\label{equ:logit}
\end{equation}
where $P(\var{Noted_{i}})$ indicates the probability of the post $i$ receiving a community note, and $\beta_{0}$ is the intercept. The vector $\bm{\var{X_{i}^{Post}}}$ contains variables describing characteristics of post content, including sentiments, topics, content types, misleadingness, and media. Additionally, the word count of post $i$ is included to control for the post length~\cite{chuai2024roll,robertson2023negativity}. Topics and media are represented as dummy variables, while all other post-level features are treated as continuous variables. The vector $\bm{\var{X_{i}^{Poster}}}$ contains variables describing characteristics of post authors, including account types, account age, followers, followees, misinformation exposure scores, and partisan scores. Account types are encoded as dummy variables, while all other author-level features are treated as continuous variables. The vectors $\bm{\beta_{1}}$ and $\bm{\beta_{2}}$ represent the coefficients of the post-level and author-level variables, respectively, and quantify their effects on the likelihood of receiving community notes. Since logistic regression coefficients are expressed in log-odds, we exponentiate them into odds ratios to facilitate interpretation and address our RQ1 and RQ2.

For RQ3, we specify a linear regression model to examine how the extracted features of community notes, posts, and post authors relate to the helpfulness and polarization of community notes:
\begin{equation}
\begin{aligned}
    Y_{j}=\, & \beta_{0} + \bm{\beta_{1}'\var{X_{j}^{Note}}} + \bm{\beta_{2}'\var{X_{j}^{Post}}} + \bm{\beta_{3}'\var{X_{j}^{Poster}}} + \epsilon_{j},
\end{aligned}
\label{equ:linear}
\end{equation}
where $Y_{j}$ represents either the helpfulness score or the polarization score of note $j$. $\beta_{0}$ is the intercept, and $\epsilon_{j}$ is the residual term for the linear regression. The vector $\bm{\var{X_{j}^{Note}}}$ includes note-specific characteristics, including domain quality and domain unbiasedness. They are treated as continuous variables. The vectors $\bm{\var{X_{j}^{Post}}}$ and $\bm{\var{X_{j}^{Poster}}}$ contains post-level and author-level features, respectively, and remain the same as defined in Eq.~\ref{equ:logit}. Coefficient estimates for $\bm{\beta_{1}}$, $\bm{\beta_{2}}$, and $\bm{\beta_{3}}$ reflect how note-, post-, and author-level attributes are associated with variations in note helpfulness and polarization, thus allowing us to assess RQ3.

\section{Empirical Results}
Here, we report our empirical results, structured around the three research questions. We begin by examining what types of requested posts are more likely to receive community notes (RQ1), then access the extent to which requests foster note writing (RQ2), and finally compare the quality of request-fostered notes with other notes in terms of helpfulness and polarization (RQ3).

\subsection{Likelihood of Receiving Community Notes}
\begin{figure*}
    \centering
    \includegraphics[width=\linewidth]{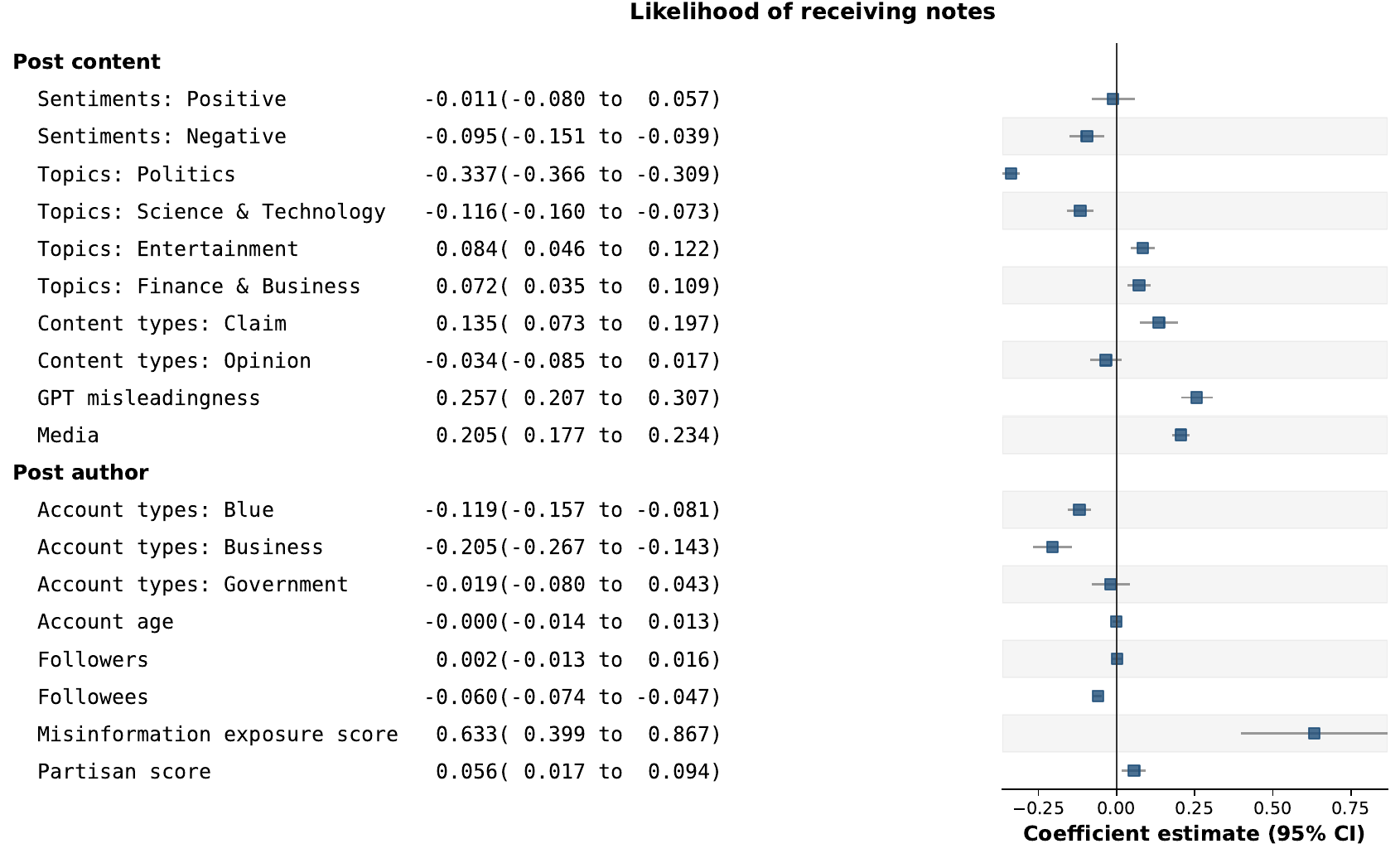}
    \caption{The estimation results for the logistic regression model predicting the likelihood that a requested post receives a community note. Shown are coefficient estimates with 95\% Confidence Intervals (CIs). The coefficients for misinformation exposure score and partisan score are estimated based on the subset of posts containing the corresponding information. The number of words in each post is controlled during estimation but omitted in visualization for better readability. Continuous independent variables---word count, account age, number of followers, and number of followees---are z-standardized before estimation to facilitate interpretation.}
    \label{fig:receive_note_coefs}
    \Description{The estimation results for the logistic regression model predicting the likelihood that a requested post receives a community note. Shown are coefficient estimates with 95\% Confidence Intervals (CIs). The coefficients for misinformation exposure score and partisan score are estimated based on the subset of posts containing the corresponding information. The number of words in each post is controlled during estimation but omitted in visualization for better readability. Continuous independent variables---word count, account age, number of followers, and number of followees---are z-standardized before estimation to facilitate interpretation.}
\end{figure*}

Of all the requested posts, \num{52862} (53.6\%) received at least one community note at the time of data collection. To investigate the factors associated with this outcome (RQ1), we estimate a logistic regression model predicting the likelihood that a requested post receives a community note. The estimation results are presented in Fig.~\ref{fig:receive_note_coefs}.

\textbf{Post content characteristics.}
We analyze how post content characteristics shape the likelihood of receiving community notes. For sentiments, we find that posts expressing stronger negative sentiment are significantly less likely to receive community notes ($\var{coef.}=-0.095$, $\var{odds~ratio}=0.909$, $p<0.01$). For instance, posts classified as purely negative exhibit 9.1\% lower odds of receiving notes compared with non-negative posts. For topics, we find that posts related to Politics ($\var{coef.}=-0.337$, $\var{odds~ratio}=0.714$, $p<0.001$) and Science \& Technology ($\var{coef.}=-0.116$, $\var{odds~ratio}=0.890$, $p<0.001$) are less likely to receive community notes compared to posts not associated with these topics. For example, political posts have 28.6\% lower odds of receiving community notes compared to non-political posts. In contrast, posts related to Entertainment ($\var{coef.}=0.084$, $\var{odds~ratio}=1.088$, $p<0.001$) and Finance \& Business ($\var{coef.}=0.072$, $\var{odds~ratio}=1.075$, $p<0.001$) are more likely to receive community notes compared to posts not associated with these topics. Additionally, posts containing media are substantially more likely to receive community notes compared to text-only posts ($\var{coef.}=0.205$, $\var{odds~ratio}=1.228$, $p<0.001$), corresponding to a 22.8\% increase in the odds of receiving community notes. For claims vs. opinions, we find that posts that contain claims are associated with 14.5\% higher odds of receiving a community note than posts without claims ($\var{coef.}=0.135$, $\var{odds~ratio}=1.145$, $p<0.001$), whereas posts expressing opinions are not significantly associated with the likelihood of receiving notes ($\var{coef.}=-0.034$, $\var{odds~ratio}=0.966$, $p=0.186$). Importantly, we find that posts with higher GPT-estimated misleadingness are significantly more likely to receive community notes ($\var{coef.}=0.257$, $\var{odds~ratio}=1.293$, $p<0.001$). Specifically, posts identified by GPT as purely misleading have 29.3\% higher odds of receiving community notes than non-misleading posts. 

\textbf{Post author characteristics.}
We further evaluate how post author characteristics shape the likelihood of receiving community notes. For verified types of post authors, we find that posts from blue ($\var{coef.}=-0.119$, $\var{odds~ratio}=0.888$, $p<0.001$) and business ($\var{coef.}=-0.205$, $\var{odds~ratio}=0.815$, $p<0.001$) accounts have 11.2\% and 18.5\% lower odds of receiving community notes than posts from unverified accounts, respectively. Additionally, posts from accounts with a higher number of followees are less likely to receive community notes, with one standard deviation increase in followers corresponding to 5.8\% decrease in the odds of receiving notes ($\var{coef.}=-0.060$, $\var{odds~ratio}=0.942$, $p<0.001$). In contrast, misinformation exposure scores are positively associated with the likelihood of receiving community notes: posts from accounts with a misinformation exposure score of 1 have 88.3\% higher odds than those from accounts with a score of 0 ($\var{coef.}=0.633$, $\var{odds~ratio}=1.883$, $p<0.001$). Moreover, posts from right-leaning accounts with a partisan score of 1 have 11.4\% higher odds of being noted compared to those from left-leaning accounts with a partisan score of \num{-1} ($\var{coef.}=0.056$, $\var{odds~ratio}=1.057$, $p<0.01$). 

In summary, the likelihood that a requested post receives a community note is shaped by both posts' and their authors' characteristics. Requested posts with negative sentiment or authored by blue/business accounts are less likely to attract notes, as are those related to politics or science. In contrast, posts related to entertainment or finance, those containing factual claims and media, and posts flagged by GPT as more misleading are significantly more likely to receive notes. At the author level, posts from accounts with a higher misinformation exposure or right-leaning partisanship (vs. left-leaning) are also more likely to be annotated. Notably, given the heightened concerns surrounding political misinformation and the fact that political posts constitute the largest part of our dataset, we additionally restrict our analysis to political posts and find that the observed patterns remain robust (see estimation results in Suppl.~\ref{sec:robustness_political}). Together, our findings highlight that Community Notes contributors who write notes and requestors who submit requests may have different selection patterns that are associated with both post features and author attributes.

\subsection{Request Timing and Request-Fostered Notes}
\label{sec:request_timing}
The request feature allows non-contributors to support the Community Notes program by flagging posts, with the goal of fostering note writing. However, receiving community notes is not necessarily a direct result of the submitted requests, as contributors can also independently select and fact-check posts regardless of requests. To assess whether the request feature meaningfully fosters note creation (RQ2), we examine the timing of requests relative to note creation.

\begin{figure*}
    \centering
    \begin{subfigure}{0.32\textwidth}
    \caption{}
    \includegraphics[width=\textwidth]{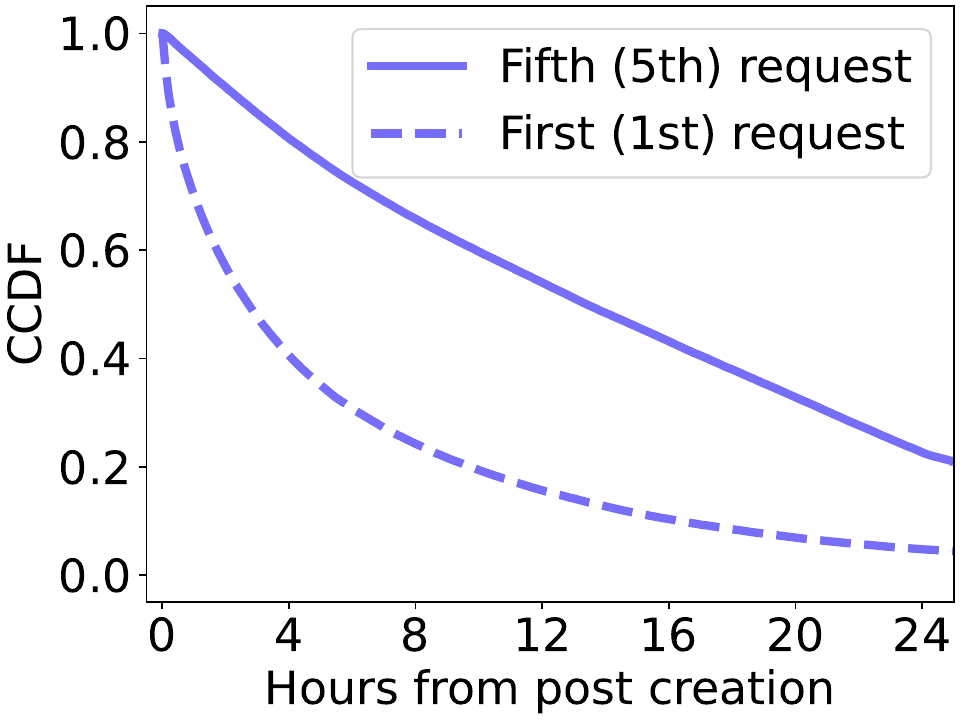}
    \label{fig:ccdf_hours2request}
    \end{subfigure}
    \hfill
    \begin{subfigure}{0.32\textwidth}
    \caption{}
    \includegraphics[width=\textwidth]{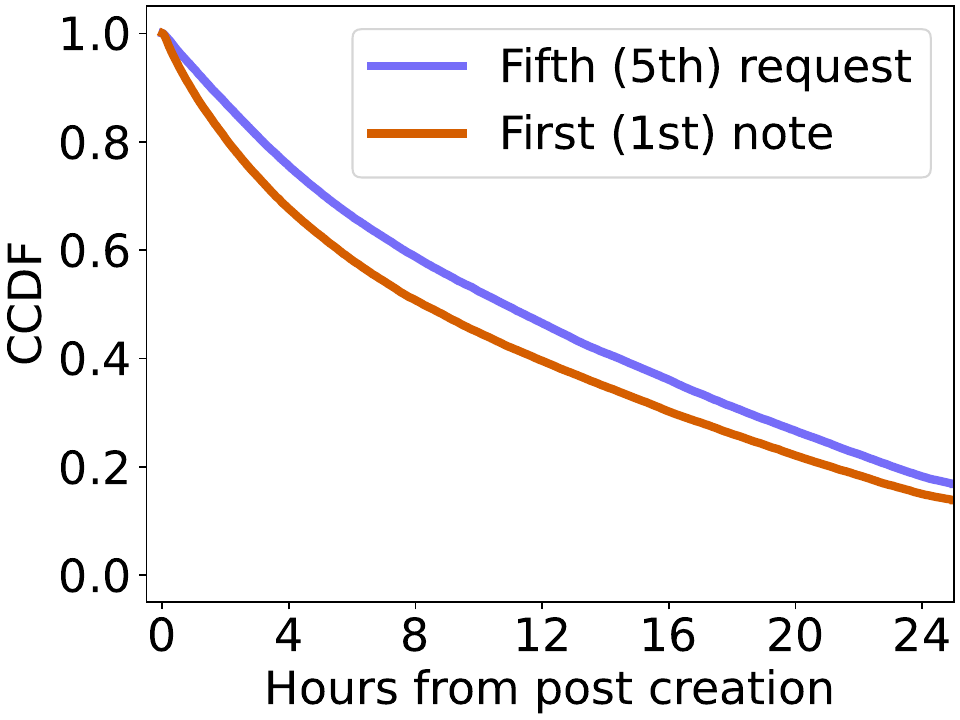}
    \label{fig:ccdf_hours2notes}
    \end{subfigure}
    \hfill
    \begin{subfigure}{0.32\textwidth}
    \caption{}
    \includegraphics[width=\textwidth]{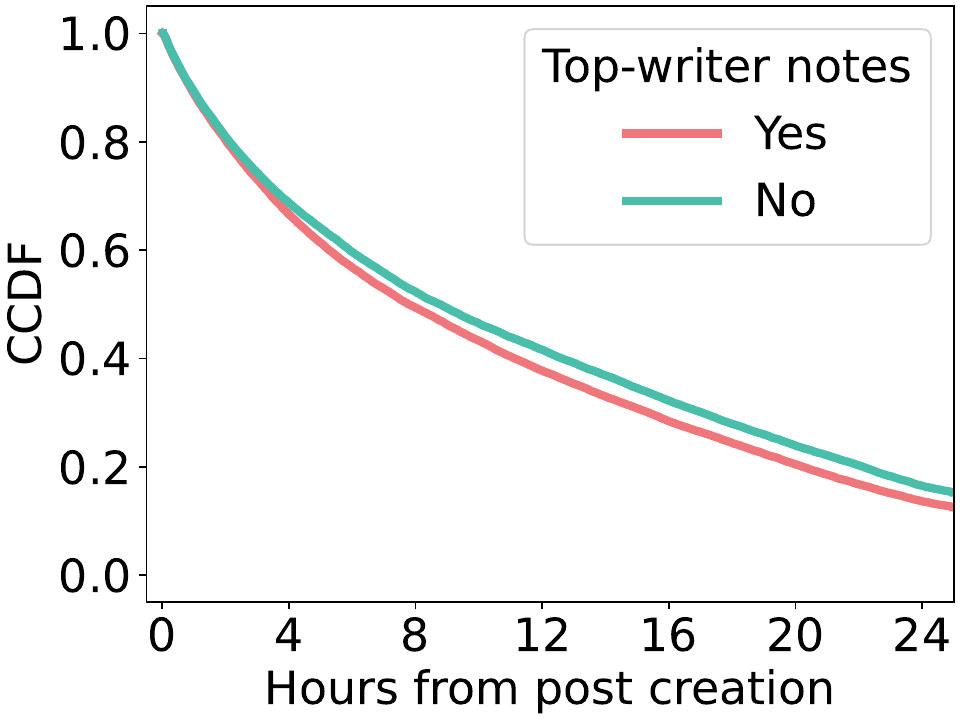}
    \label{fig:ccdf_top_writers}
    \end{subfigure}

    \begin{subfigure}{\textwidth}
    \caption{}
    \includegraphics[width=\textwidth]{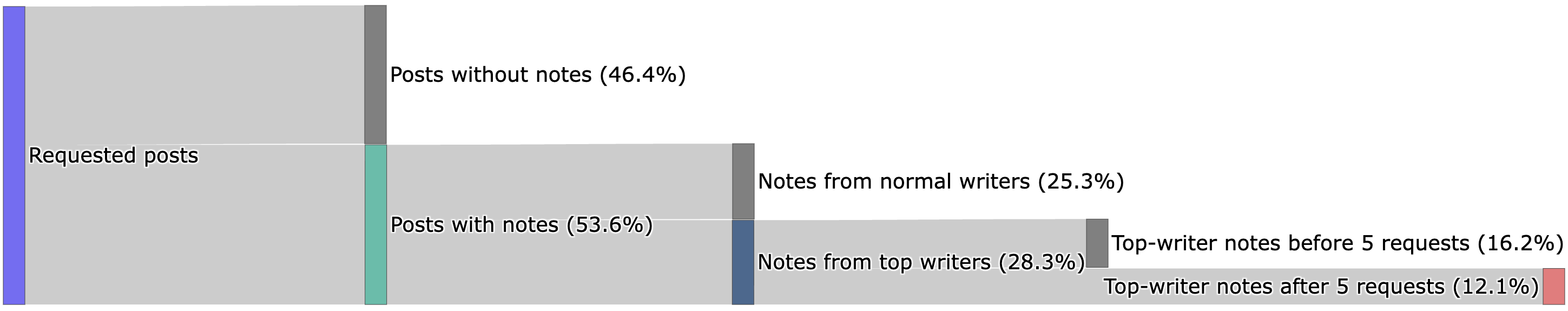}
    \label{fig:sankey_posts}
    \end{subfigure}
    \caption{The statistics for the request timing relative note writing. (a)~The Complementary Cumulative Distribution Functions (CCDFs) for the hours from post creation to the submission of first request and to the submission of fifth request. (b)~The CCDFs for the hours from post creation to the submission of fifth request and to the generation of first community note, respectively. (c)~The CCDFs for the hours from post creation to the notes written by top writers versus other contributors. (d)~The sankey plot illustrating the estimated proportion of posts for which community notes are likely fostered by requests.}
    \label{fig:requests2notes}
    \Description{The statistics for the request timing relative note writing. (a)~The Complementary Cumulative Distribution Functions (CCDFs) for the hours from post creation to the submission of first request and to the submission of fifth request. (b)~The CCDFs for the hours from post creation to the submission of fifth request and to the generation of first community note, respectively. (c)~The CCDFs for the hours from post creation to the notes written by top writers versus other contributors. (d)~The sankey plot illustrating the estimated proportion of posts for which community notes are likely fostered by requests.}
\end{figure*}

\textbf{Request timing relative to note creation.}
As shown in Fig.~\ref{fig:ccdf_hours2request}, the median time from post creation to the submission of first request is 2.72 hours, while it takes significantly longer for a post to accumulate five requests (median of 13.40 hours; $p_{MWU}<0.001$). We find that the response time of community notes (median of 8.23 hours) is significantly shorter than the time it typically takes for posts to receive five requests (Fig.~\ref{fig:ccdf_hours2notes}; $p_{MWU}<0.001$). Furthermore, Fig.~\ref{fig:ccdf_top_writers} shows that top writers (median of 7.78 hours) generate notes almost an hour faster than other contributors (median of 8.76 hours; $p_{MWU}<0.001$). Taken together, these findings suggest that community notes tend to be written before the fifth request is submitted (\ie, request threshold)---implying that requests may not be the primary driver of note creation. Additionally, because community notes often appear more quickly than request alerts, and because top writers tend to response faster than other contributors, notes generated by top writers after the fifth request to the corresponding posts are highly likely to have been shaped by request alerts. Therefore, we consider top-writer notes generated after the request threshold as request-fostered notes. Consistent with the guidelines of the request feature, we find that only an estimated 12.1\% of requested posts were likely influenced by the request activity, as their associated notes were written by top writers after the request threshold was reached (Fig.~\ref{fig:sankey_posts}). Therefore, the effect of the request feature in driving note creation is still limited.

\begin{figure*}
    \centering
    \includegraphics[width=\linewidth]{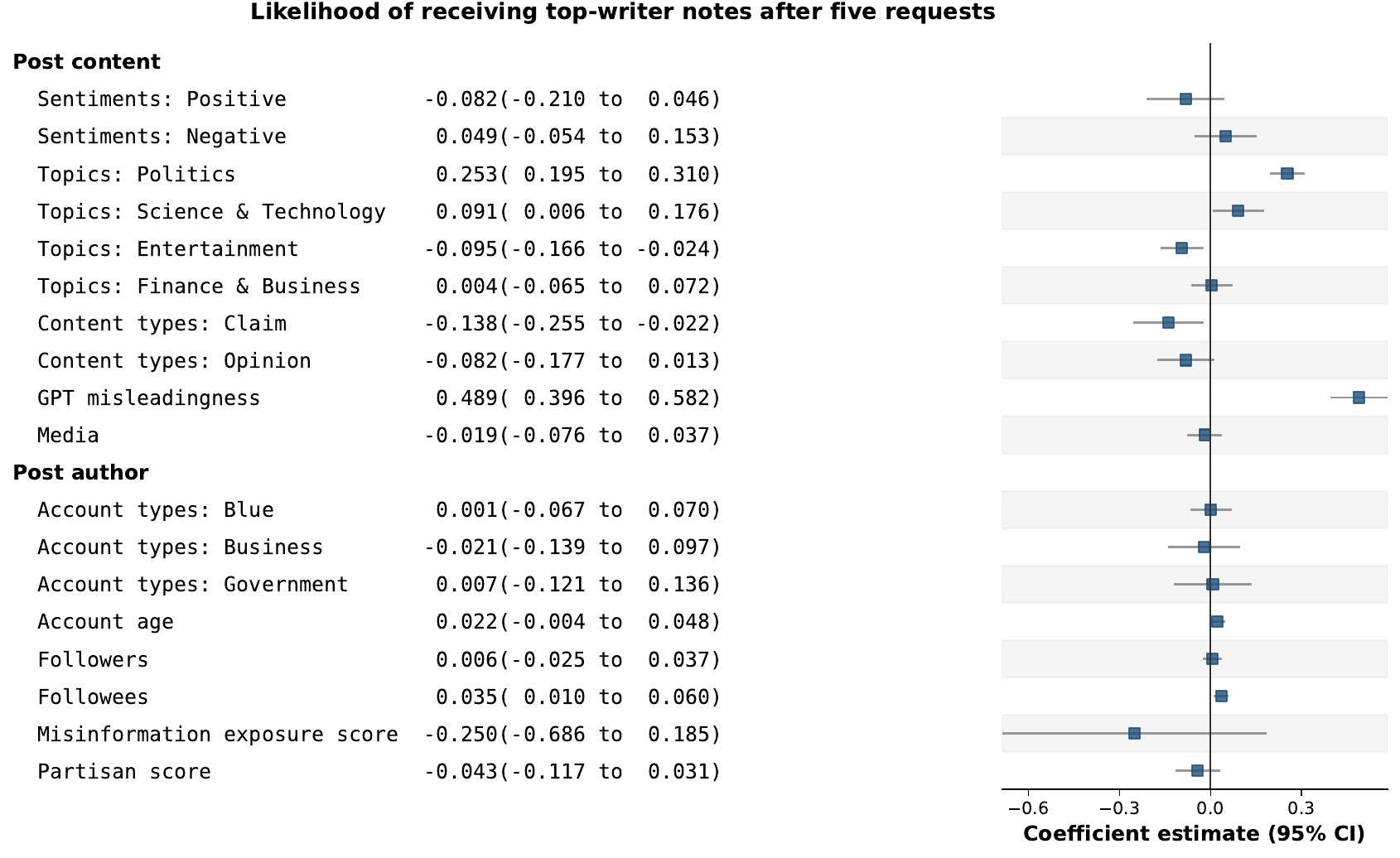}
    \caption{The estimation results for the logistic regression model predicting the likelihood that a requested post receives a top-writer note after the request threshold, \ie, request-fostered notes. Shown are coefficient estimates with 95\% CIs. The coefficients for misinformation exposure score and partisan score are estimated based on the subset of posts containing the corresponding information. The number of words in each post is controlled during estimation but omitted in visualization for better readability. Continuous independent variables---word count, account age, number of followers, and number of followees---are z-standardized before estimation to facilitate interpretation.}
    \label{fig:top_writer_note_coefs}
    \Description{The estimation results for the logistic regression model predicting the likelihood that a requested post receives a top-writer note after the request threshold, i.e., request-fostered notes. Shown are coefficient estimates with 95\% CIs. The coefficients for misinformation exposure score and partisan score are estimated based on the subset of posts containing the corresponding information. The number of words in each post is controlled during estimation but omitted in visualization for better readability. Continuous independent variables---word count, account age, number of followers, and number of followees---are z-standardized before estimation to facilitate interpretation.}
\end{figure*}

\textbf{Request-fostered notes from top writers.}
We further examine whether the request function influences the fact-checking behavior of top writers. To this end, we employ the logistic regression model as specified in Eq.~\ref{equ:logit} and compare top writers' selection patterns when exposed to request alerts versus when writing notes without such alerts. Here, the dependent variable is the probability that a post receives a request-fostered note. Under the null hypothesis that requests do not affect selection behavior of top writers, the coefficient estimates for all independent variables should be statistically indistinguishable from zero; statistically significant coefficient estimates, by contrast, indicate systematic differences in post selection following request exposure. Importantly, to ensure a fair comparison, we restrict our analysis to posts on which notes were generated only by topic writers before and after the request threshold. The estimation results are presented in Fig.~\ref{fig:top_writer_note_coefs}. 

We find that top-writer notes on posts related to Politics ($\var{coef.}=0.253$, $\var{odds~ratio}=1.287$, $p<0.001$) and Science \& Technology ($\var{coef.}=0.091$, $\var{odds~ratio}=1.095$, $p<0.05$) have 28.7\% and 9.5\% higher odds of being request-fostered, respectively, compared to posts without the corresponding topics. This suggests that top writers select more political and scientific posts when receiving request alerts. In contrast, top-writer notes on posts related to Entertainment ($\var{coef.}=-0.095$, $\var{odds~ratio}=0.910$, $p<0.01$) and containing claims ($\var{coef.}=-0.138$, $\var{odds~ratio}=0.871$, $p<0.05$) have 9\% and 12.9\% lower odds of being request-fostered, respectively. This indicates that top writers choose less entertainment-related and claim-oriented content after receiving request alerts, compared to without receiving alerts. Together, these results suggest that requests can shift contributors' attention toward political and scientific content, while diverting attention away from claim-oriented posts or those related to entertainment. In particular, the coefficient estimate for GPT-estimated misleadingness is positive and substantially significant ($\var{coef.}=0.489$, $\var{odds~ratio}=1.631$, $p<0.001$). Posts flagged by GPT as misleading (score $=1$) have 63.1\% higher odds of receiving request-fostered top-writer notes than non-misleading posts (score $=0$). This suggests that top writers prioritize posts with higher potential misleadingness when guided by request alerts, compared to when no alerts are present.

In summary, out of all requested posts eligible for community evaluation, only 12.1\% appear to have been generated by top writers through request alerts. Furthermore, requests typically arrive later than community notes themselves, as the time from post creation to request alerts is significantly longer than the time from post creation to note generation, limiting the request feature's potential to foster note writing. Nevertheless, by comparing notes written by top writers before and after the request threshold, we find that requests shift contributors' attention toward political content while prioritizing posts with higher potential misleadingness.\footnote{To further reduce unobserved confounding at the contributor level, including heterogeneity among top writers, we additionally restrict the analysis to posts selected by the same set of top writers before and after the request threshold (see details in Suppl.~\ref{sec:robustness_top_writers}). This conservative approach confirms that our main findings remain robust, particularly with respect to political content and posts with high potential misleadingness.} Thus, although the overall effect of requests on note creation remains modest, the feature shows promise in channeling contributor attention toward high-stakes content at greater risk of misinformation.

\subsection{Algorithmic Evaluation of Community Notes}
\begin{figure*}
    \centering
    \begin{subfigure}{0.32\textwidth}
    \caption{}
    \includegraphics[width=\textwidth]{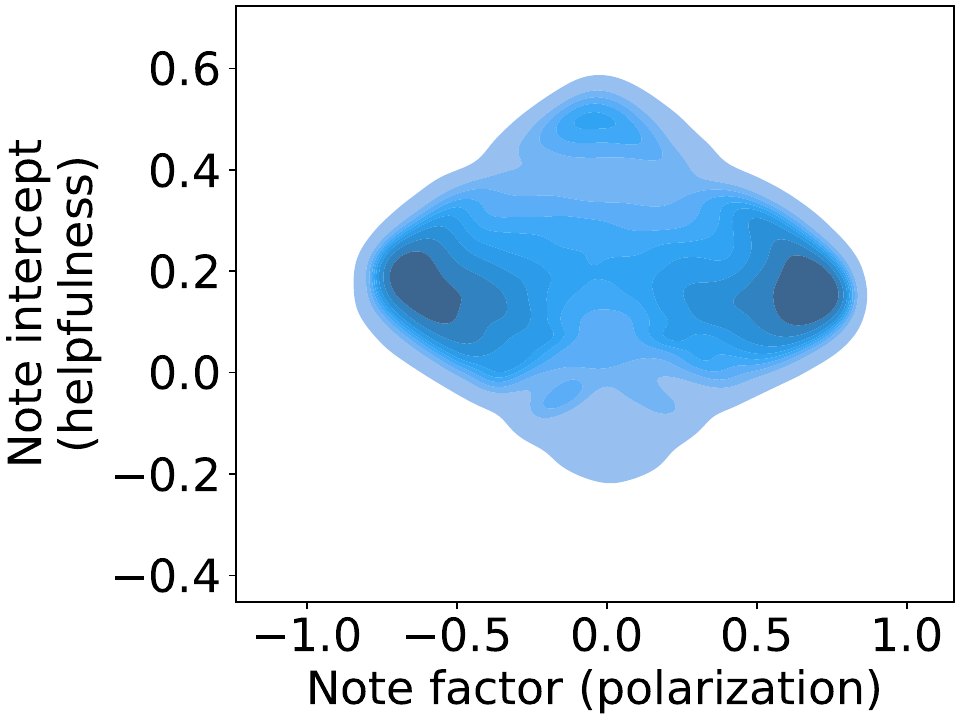}
    \label{fig:note_eval_original}
    \end{subfigure}
    \hfill
    \begin{subfigure}{0.32\textwidth}
    \caption{}
    \includegraphics[width=\textwidth]{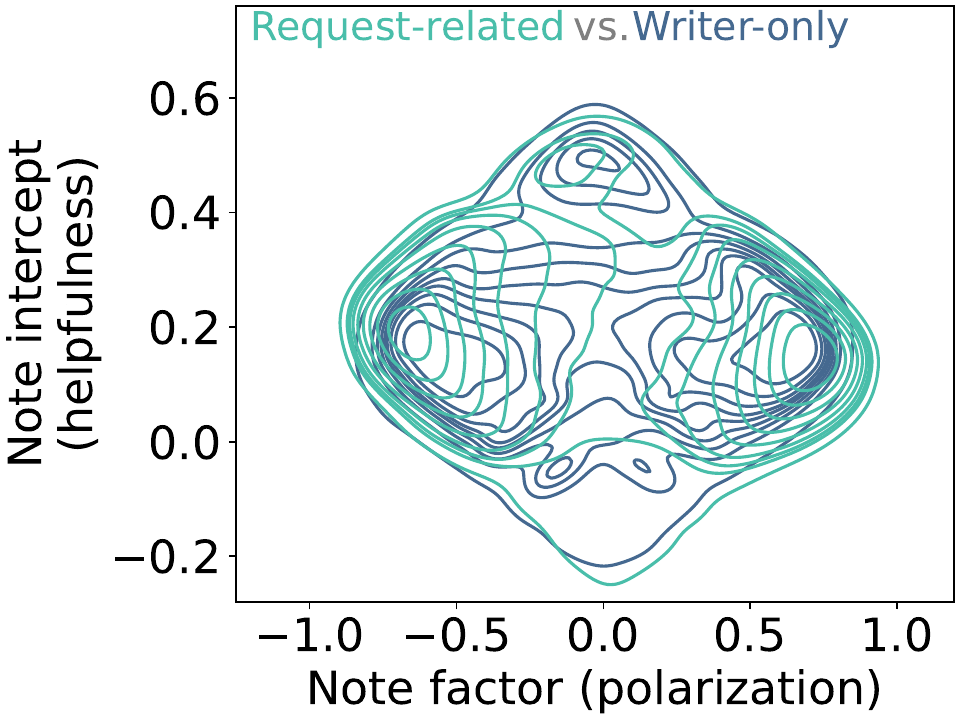}
    \label{fig:note_eval_request}
    \end{subfigure}
    \hfill
    \begin{subfigure}{0.32\textwidth}
    \caption{}
    \includegraphics[width=\textwidth]{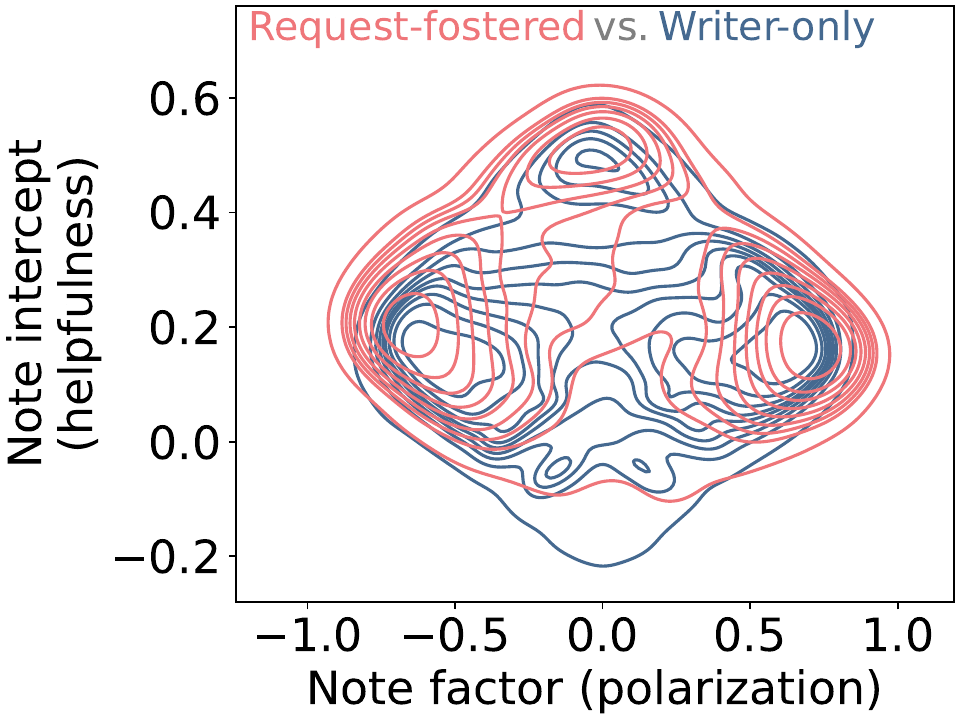}
    \label{fig:note_eval_request_fostered}
    \end{subfigure}

    \begin{subfigure}{0.32\textwidth}
    \caption{}
    \includegraphics[width=\textwidth]{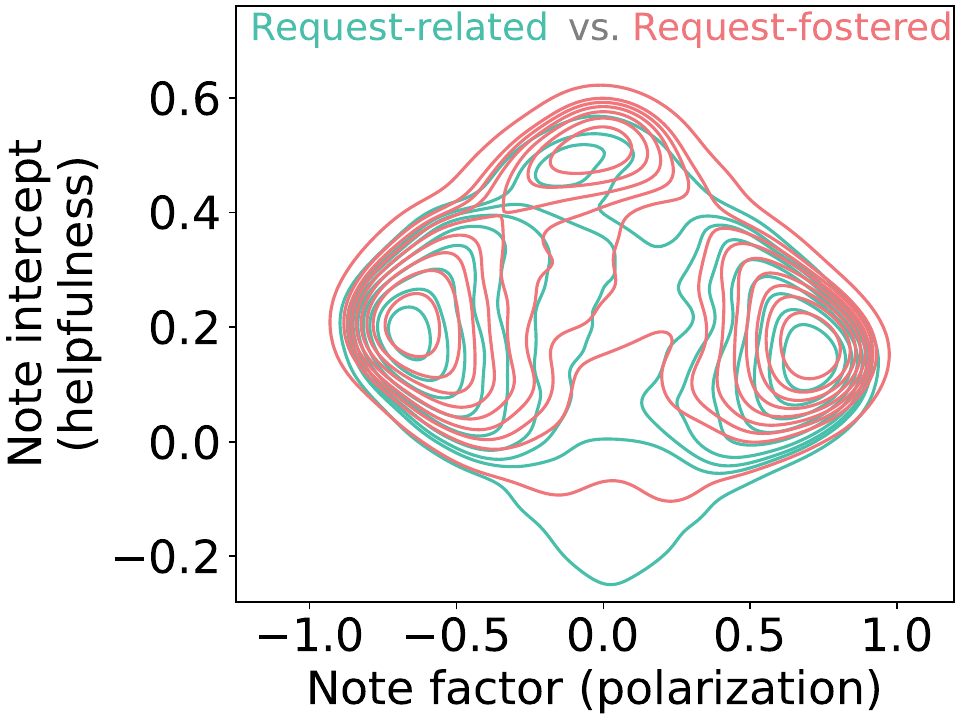}
    \label{fig:note_eval_after_request}
    \end{subfigure}
    \hfill
    \begin{subfigure}{0.32\textwidth}
    \caption{}
    \includegraphics[width=\textwidth]{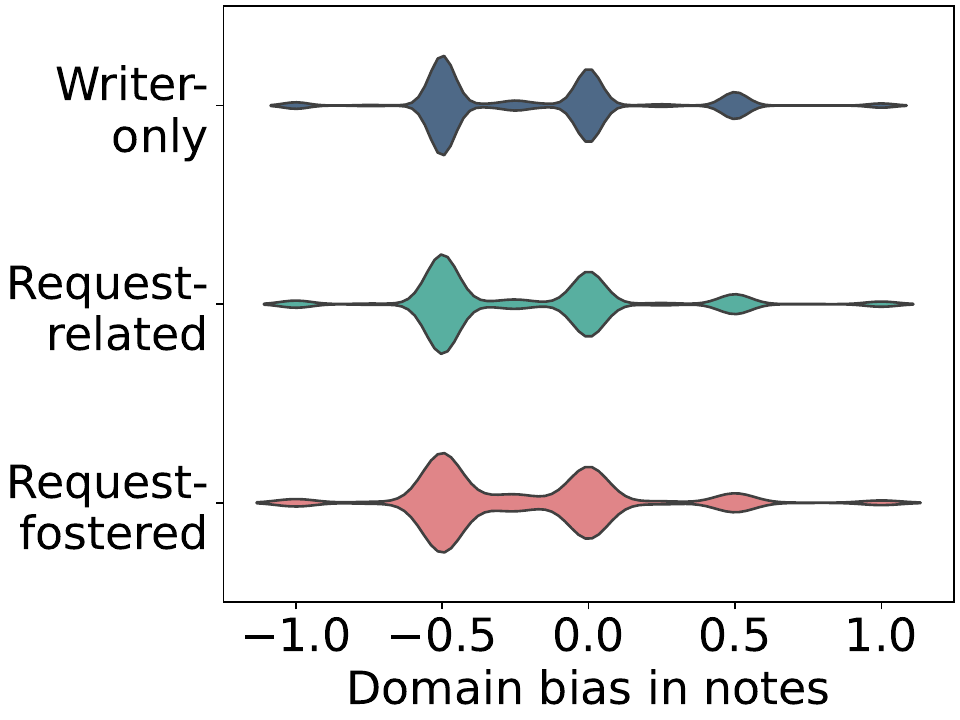}
    \label{fig:domain_bias}
    \end{subfigure}
    \hfill
    \begin{subfigure}{0.32\textwidth}
    \caption{}
    \includegraphics[width=\textwidth]{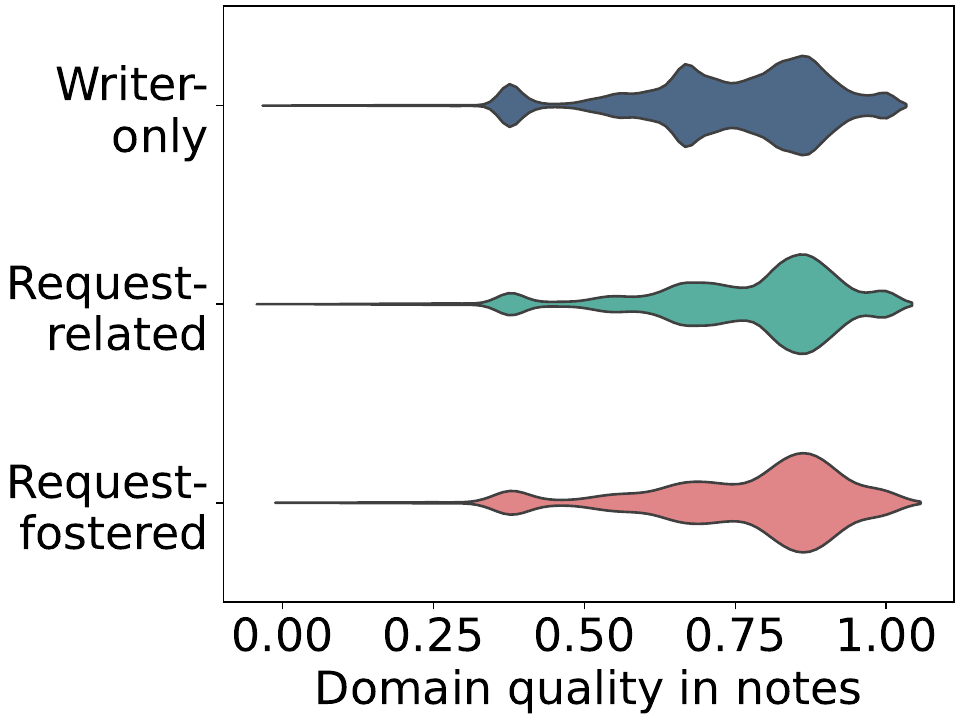}
    \label{fig:domain_quality}
    \end{subfigure}
    \caption{Overview of note evaluations and source domains in community notes. (a)~The distribution of estimated note intercepts (helpfulness) and estimated note factors (polarization) from the note selection algorithm. (b)~The distributions of estimated note intercepts (helpfulness) and note factors (polarization) from the note selection algorithm between request-related notes and writer-only notes. (c)~The distributions of estimated note intercepts (helpfulness) and note factors (polarization) from the note selection algorithm between request-fostered notes and writer-only notes. (d)~The distributions of estimated note intercepts (helpfulness) and note factors (polarization) from the note selection algorithm between request-related notes and request-fostered notes. (e)~The violin plots showing the distributions of domain bias in community notes across the three categories: writer-only, request-related, and request-fostered. (f)~The violin plots showing the distributions of domain quality in community notes across the three categories: writer-only, request-related, and request-fostered.}
    \label{fig:overview_note}
    \Description{Overview of note evaluations and source domains in community notes. (a)~The distribution of estimated note intercepts (helpfulness) and estimated note factors (polarization) from the note selection algorithm. (b)~The distributions of estimated note intercepts (helpfulness) and note factors (polarization) from the note selection algorithm between request-related notes and writer-only notes. (c)~The distributions of estimated note intercepts (helpfulness) and note factors (polarization) from the note selection algorithm between request-fostered notes and writer-only notes. (d)~The distributions of estimated note intercepts (helpfulness) and note factors (polarization) from the note selection algorithm between request-related notes and request-fostered notes. (e)~The violin plots showing the distributions of domain bias in community notes across the three categories: writer-only, request-related, and request-fostered. (f)~The violin plots showing the distributions of domain quality in community notes across the three categories: writer-only, request-related, and request-fostered.}
\end{figure*}

To address RQ3, we downloaded and rerun the source code of the note selection algorithm released by \X. Fig.~\ref{fig:note_eval_original} presents the distribution of note intercepts and note factors from the note selection algorithm. The note intercepts reflect the perceived helpfulness of community notes across raters, with higher intercepts indicating higher helpfulness. The note factors capture the polarization in ratings, with values deviating from 0 indicating greater disagreement among raters regarding the helpfulness of community notes. Here, we categorize all community notes into three groups: \emph{writer-only} (notes on posts that received fewer than five requests and were solely written by contributors), \emph{request-fostered} (notes authored by top writers after the fifth request, likely influenced by requests), and \emph{request-related} (all remaining notes on posts that received five or more requests but not classified as ``request-fostered''). Notably, request-related notes may also be written by top writers; however, unlike request-fostered notes, they were produced before the threshold of request alerts. We then analyze the helpfulness and polarization (measured as the absolute value of note factors) of community notes across these categories.

\textbf{Helpfulness and polarization of request-fostered notes.}
We examine the distributions of note intercepts and note factors across writer-only notes, request-related notes, and request-fostered notes.
\begin{itemize}[leftmargin=*]
    \item Request-related notes vs. writer-only notes (Fig.~\ref{fig:note_eval_request}): Request-related notes exhibit lower helpfulness (mean $=$ 0.172 vs. 0.178; $p_{MWU}<0.001$) and greater polarization (mean $=$ 0.453 vs. 0.343; $p_{MWU}<0.001$), compared to writer-only notes.
    \item Request-fostered notes vs. writer-only notes (Fig.~\ref{fig:note_eval_request_fostered}): Request-fostered notes are rated as more helpful (mean $=$ 0.231 vs. 0.178; $p_{MWU}<0.001$) but also more polarized (mean $=$ 0.424 vs. 0.343; $p_{MWU}<0.001$) than writer-only notes.
    \item Request-related notes vs. request-fostered notes (Fig.~\ref{fig:note_eval_after_request}): Request-fostered notes are rated as more helpful (mean $=$ 0.231 vs. 0.172; $p_{MWU}<0.001$) and less polarized (mean $=$ 0.424 vs. 0.453; $p_{MWU}<0.001$) than request-related notes.
\end{itemize}
Taken together, these findings indicate that, although requests in general are associated with less helpful and more polarized notes, request-fostered notes written by top writers tend to achieve higher quality---being both more helpful and less polarized than request-related notes. To further substantiate this pattern, we estimate a logistic regression predicting whether a note is request-fostered versus request-related using helpfulness and polarization as predictors. The results show a strong positive association for helpfulness ($\var{coef.}=2.528$, $\var{odds~ratio}=12.530$, $p<0.001$) and a significant negative association for polarization ($\var{coef.}=-0.231$, $\var{odds~ratio}=0.794$, $p<0.001$), confirming that request-fostered notes stand out within the request context by being both more helpful and less polarized than other notes. Subsequently, we investigate what factors could explain this quality advantage.

\begin{figure*}
    \centering
    \includegraphics[width=\linewidth]{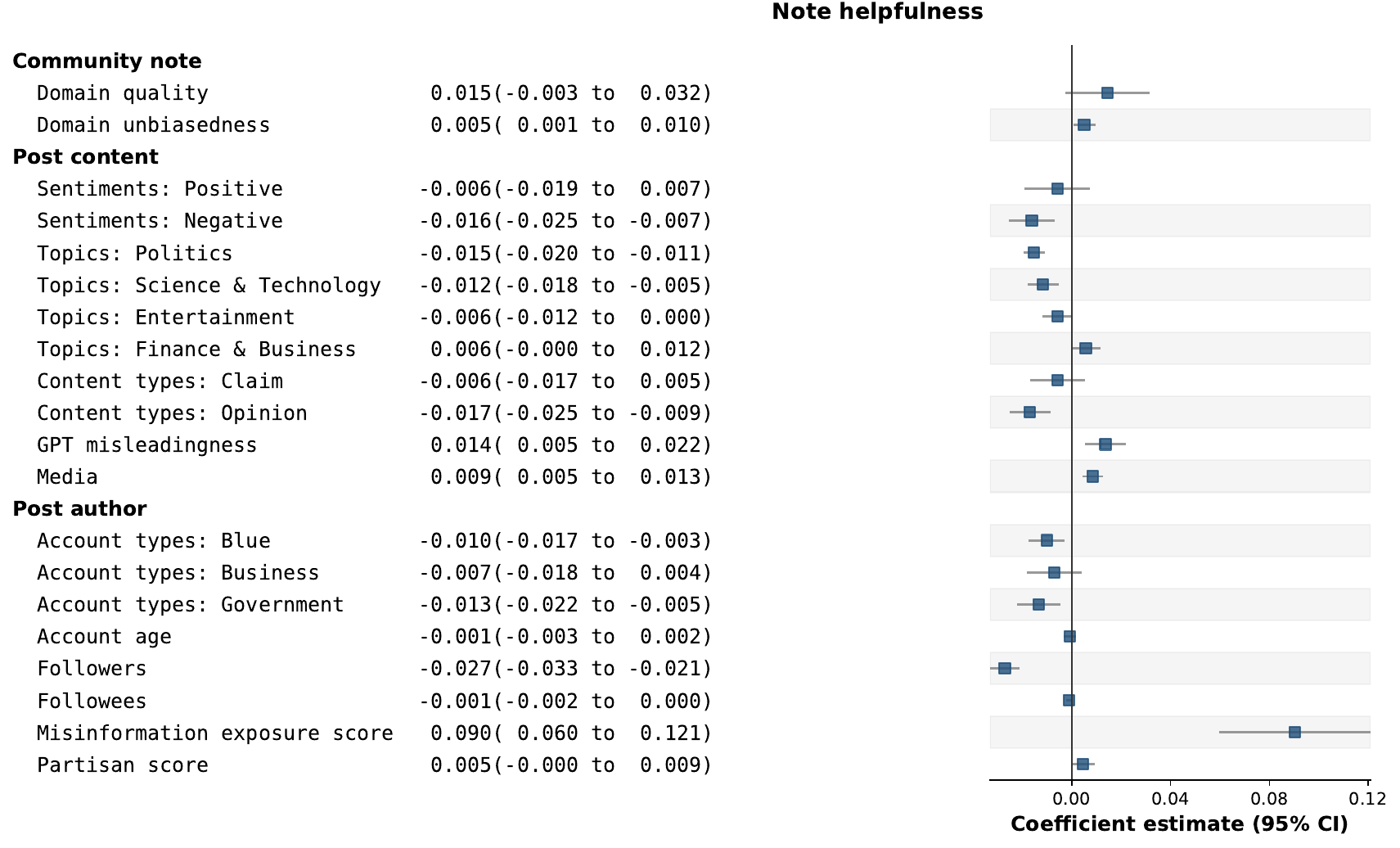}
    \caption{The estimation results for the linear regression model predicting note helpfulness. Shown are coefficient estimates with 95\% CIs. The number of words in each post is controlled during estimation but omitted in visualization for better readability. Continuous independent variables---word count, account age, number of followers, and number of followees---are z-standardized before estimation to facilitate interpretation.}
    \label{fig:note_score_coefs}
    \Description{The estimation results for the linear regression model predicting note helpfulness. Shown are coefficient estimates with 95\% CIs. The number of words in each post is controlled during estimation but omitted in visualization for better readability. Continuous independent variables---word count, account age, number of followers, and number of followees---are z-standardized before estimation to facilitate interpretation.}
\end{figure*}

Since contributors are required to cite external sources when writing community notes, we evaluate cited domains in terms of political bias and information quality to probe possible explanations for the higher quality of request-fostered notes than other notes. Overall, we find that community notes tend to cite more left-leaning domains than right-leaning ones (Fig.~\ref{fig:domain_bias}). Domains in request-related (mean of \num{-0.210}) are slightly more left-leaning than those in writer-only notes (mean of \num{-0.185}; $p_{MWU}<0.001$). The political bias of domains in request-fostered notes (mean of \num{-0.214}) has no statistically significant difference from request-related notes ($p_{MWU}=0.352$). In terms of information quality (Fig.~\ref{fig:domain_quality}), request-related notes cite higher-quality domains (mean of \num{0.769}) than writer-only notes (mean of \num{0.742}; $p_{MWU}<0.001$), whereas request-fostered notes (mean of \num{0.767}) do not differ from request-related notes ($p_{MWU}=0.834$). These findings suggest that the higher quality of request-fostered notes is unlikely to be explained by domain choices alone.

Because Community Notes contributors are anonymized within the system, rater bias and polarization toward specific writers' identities (\eg, partisanship) are reduced~\cite{cn2025signing}. Consequently, observed differences in helpfulness or polarization are more likely to reflect the types of posts and authors contributors choose to fact-check. We therefore examine which notes, posts, and authors are associated with higher helpfulness or polarization.

\textbf{Factors associated with note helpfulness.}
The estimation results from the linear regression model for note helpfulness (as specified in Eq.~\ref{equ:linear}) are shown in Fig.~\ref{fig:note_score_coefs}. We find that the unbiasedness (\ie, neither left-leaning nor right-leaning) of external domains cited in community notes is positively associated with note helpfulness ($\var{coef.}=0.005$, $p<0.05$), suggesting that notes referencing politically neutral sources are perceived as more helpful compared to those citing left- or right-leaning domains. In contrast, the information quality of external domains is not significantly associated with helpfulness ($\var{coef.}=0.015$, $p=0.095$).

Regarding post content, notes on posts with higher negative sentiment are rated as less helpful ($\var{coef.}=-0.016$, $p<0.01$). Posts related to Politics ($\var{coef.}=-0.015$, $p<0.001$) and Science \& Technology ($\var{coef.}=-0.012$, $p<0.001$) tend to receive less-helpful notes compared to posts without these topics. Additionally, posts with higher opinion scores are associated with lower note helpfulness ($\var{coef.}=-0.017$, $p<0.001$). Conversely, community notes on posts containing media ($\var{coef.}=0.009$, $p<0.001$) or with higher estimated misleadingness ($\var{coef.}=0.014$, $p<0.01$) are rated as more helpful. With respect to post authors, notes on posts from the blue ($\var{coef.}=-0.010$, $p<0.01$) and government ($\var{coef.}=-0.013$, $p<0.01$) accounts, as well as from authors with more followers ($\var{coef.}=-0.027$, $p<0.001$), are less helpful compared to those from unverified accounts with fewer followers. In contrast, higher misinformation exposure scores of post authors are positively associated with note helpfulness ($\var{coef.}=0.090$, $p<0.001$).

In summary, the helpfulness of community notes is positively associated with (i) citations to unbiased domains, (ii) posts containing media, (iii) post misleadingness, and (iv) misinformation exposure scores of post authors. Conversely, the helpfulness of community notes is negatively associated with (i) negative sentiment in posts, (ii) political and scientific topics, (iii) opinionated content, (iv) blue and government accounts, and (v) the number of followers.

\begin{figure*}
    \centering
    \includegraphics[width=\linewidth]{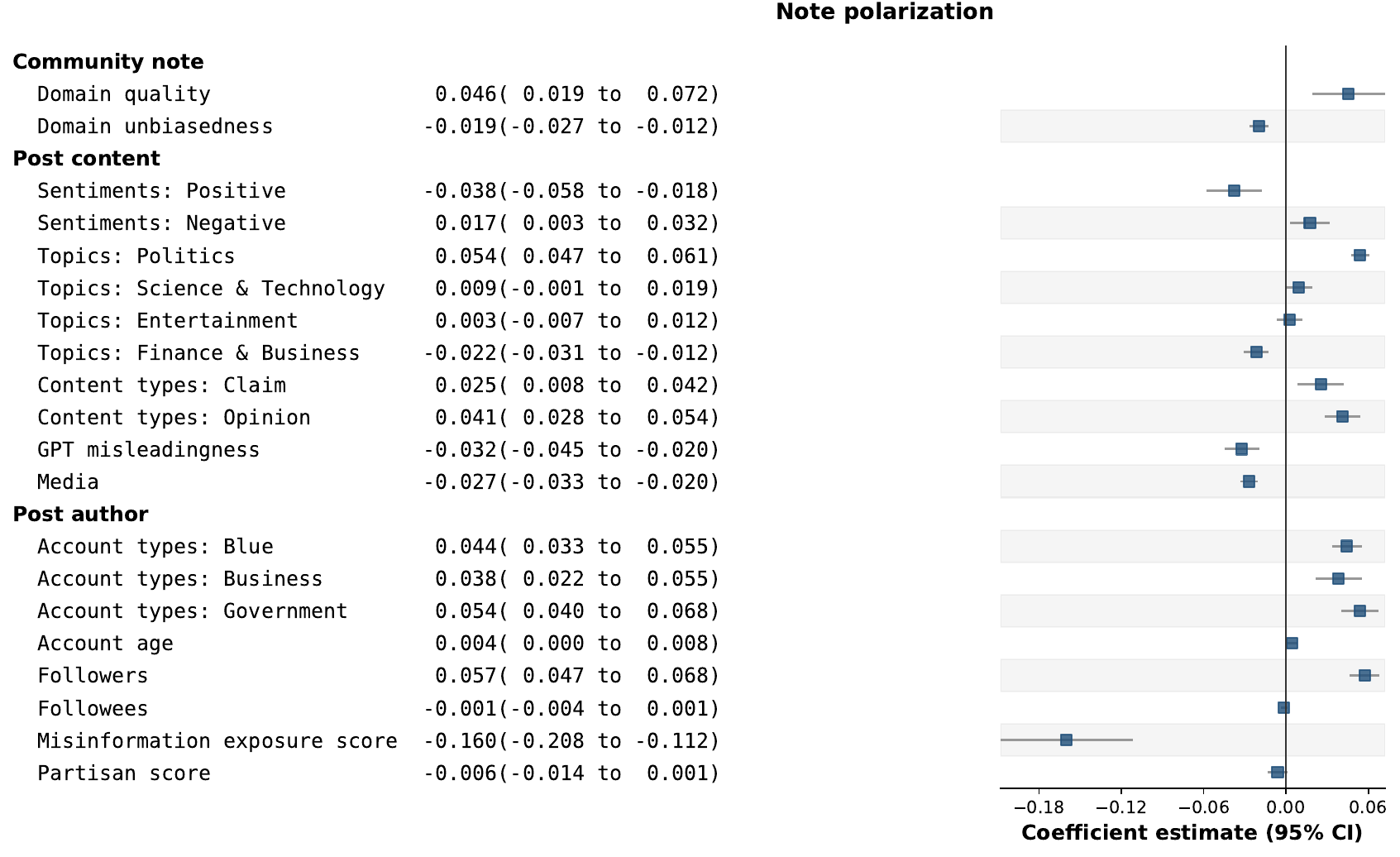}
    \caption{The estimation results for the linear regression model predicting note polarization. Shown are coefficient estimates with 95\% CIs. The number of words in each post is controlled during estimation but omitted in visualization for better readability. Continuous independent variables---word count, account age, number of followers, and number of followees---are z-standardized before estimation to facilitate interpretation.}
    \label{fig:note_factor_coefs}
    \Description{The estimation results for the linear regression model predicting note polarization. Shown are coefficient estimates with 95\% CIs. The number of words in each post is controlled during estimation but omitted in visualization for better readability. Continuous independent variables---word count, account age, number of followers, and number of followees---are z-standardized before estimation to facilitate interpretation.}
\end{figure*}

\textbf{Factors associated with note polarization.}
The estimation results for a linear regression model predicting note polarization are shown in Fig.~\ref{fig:note_factor_coefs}. Community notes that cite politically unbiased domains tend to have reduced polarization across their ratings ($\var{coef.}=-0.019$, $p<0.001$), whereas citations to domains with higher information quality are associated with increased polarization ($\var{coef.}=0.046$, $p<0.01$).

In terms of post content, positive sentiment in posts is negatively associated with note polarization ($\var{coef.}=-0.038$, $p<0.001$), but negative sentiment shows the opposite pattern ($\var{coef.}=0.017$, $p<0.05$). Posts related to Politics are linked to higher polarization ($\var{coef.}=0.054$, $p<0.001$), whereas posts related to Finance \& Business ($\var{coef.}=-0.022$, $p<0.001$) are associated with lower polarization. Community notes on posts containing stronger claims ($\var{coef.}=0.025$, $p<0.001$) or more opinionated content ($\var{coef.}=0.041$, $p<0.001$) tend to be more polarized. In contrast, community notes on posts that include media ($\var{coef.}=-0.027$, $p<0.001$) or exhibit higher misleadingness ($\var{coef.}=-0.032$, $p<0.001$) have lower polarization scores. For post authors, posts from all types of verified accounts---blue ($\var{coef.}=0.044$, $p<0.001$), business ($\var{coef.}=0.038$, $p<0.001$), and government ($\var{coef.}=0.054$, $p<0.001$)---tend to receive community notes with higher polarization compared to posts from unverified accounts. Additionally, posts authored by accounts with more followers are associated with greater note polarization ($\var{coef.}=0.057$, $p<0.001$), and posts from accounts with higher misinformation exposure scores are more likely to receive community notes with reduced polarization ($\var{coef.}=-0.160$, $p<0.001$).

In summary, the polarization of community notes is positively associated with (i) high quality of external domains, (ii) negative sentiment, (iii) political topics, (iv) posts containing claims or opinionated content, and (v) verified accounts with many followers. Conversely, the polarization of community notes is negatively associated with (i) citations to unbiased domains, (ii) positive sentiment, (iii) posts containing media, (iv) misleadingness of post content and (v) misinformation exposure scores of post authors.

Notably, through the analysis of both note helpfulness and polarization, we observe that post misleadingness is positively associated with note helpfulness but negatively associated with note polarization, whereas political content exhibits the opposite pattern. Together with our findings in Section \ref{sec:request_timing}, which show that requests shift top writers' attention toward political content while prioritizing posts with higher potential misleadingness, our analysis provides additional insight: the higher helpfulness and lower polarization of request-fostered notes can be partially interpreted as reflecting top writers' selective fact-checking of misleading posts.

\subsection{Summary of Main Findings}
In this study, based on the request feature implemented on \X and associated requests, posts, and community notes, we conduct a comprehensive analysis on the Community Notes system with respect to content selection, contributor behavior, and note evaluation. Overall, our analysis highlights three key findings.
\begin{itemize}[leftmargin=*]
    \item Requested posts are more likely to receive notes when they are associated with entertainment, finance, media, claims, high misleadingness, or authors with high misinformation exposure, but less so for negative, political or scientific content. This suggests that Community Notes contributors and requestors have distinct selection patterns. 
    \item Only an estimated 12.1\% of the requested posts appear to receive request-fostered notes from top writers, with many notes written independently of the request feature. Nevertheless, requests can shape contributors' behavior by directing their focus toward politically salient content, while still prioritizing posts with higher risk of misinformation.
    \item Request-fostered notes exhibit higher helpfulness and lower polarization compared to notes generated prior to the request threshold or by non–top writers, a pattern that may partly reflect top writers selectively fact-checking posts with higher misleadingness.
\end{itemize}

\section{Discussion}
Although the Community Notes system has emerged as a promising model of crowdsourced fact-checking, its speed and coverage remain limited, constraining its scalability on social media platforms~\cite{de2025supernotes,chuai2024community.new}. To address these challenges, \X recently introduced the Request Community Note feature, which opens a new channel for fact-checking participation and enables a broader set of users to actively solicit notes on specific posts. In this study, we analyze a large-scale dataset of \num{98685} posts surfaced through the request feature, along with their associated notes, to examine how this feature influences \X's Community Notes system. Our findings shed light on the role of requests in shaping what content is fact-checked, how contributors engage, and the quality of resulting notes.

\subsection{Research Implications}
We find that more than half of the requested posts (53.6\%) ultimately receive community notes. Yet, the processes of submitting requests and generating notes are largely independent, with only 12.1\% of posts receiving annotations that can be potentially attributed to the request feature. This suggests that requests do not strongly determine contributor behavior. Instead, the relative independence of the two processes makes requests a useful lens for examining how Community Notes contributors select posts for fact-checking and how their choices align---or diverge---from the priorities expressed by users who submit requests (\ie, requestors). 

\textbf{Likelihood of receiving community notes.}
Our results show that the likelihood of receiving community notes for requested posts are shaped by both post content and author characteristics. Specifically, posts with negative sentiment, authored by blue or business accounts, or focused on politics or science are less likely to receive community notes. In contrast, posts related to entertainment or finance, those containing claims or media, and those flagged by GPT as more misleading are significantly more likely to receive community notes. At the post author level, posts from accounts with higher misinformation exposure or right-leaning partisanship are also more likely to be annotated. The misalignment between the types of posts surfaced by requestors (\eg, political content) and those annotated by contributors potentially underscores a divergence in fact-checking selection.\footnote{To validate our findings on content selection by contributors and rule out confounding from request-driven activity, we exclude posts with request-fostered community notes (as identified in Section~\ref{sec:request_timing}) and repeat our analysis. The results remain robust (see details in Suppl.~\ref{sec:robustness_likelihood}).}

On the one hand, claim-oriented posts with higher estimated misleadingness and from authors with higher misinformation exposure scores are more likely to receive community notes. This suggests that, compared to requestors, contributors might adopt stricter standards, prioritizing content with greater potential for misinformation and harm. On the other hand, previous research finds that contributors often fact-check posts that are relatively straightforward to verify or contain claims already addressed in earlier expert fact-checks~\cite{borenstein2025community}. Similarly, LLMs perform strongly when processing logically structured claims but struggle with more complex ones~\cite{costabile2025assessing}. This suggests that, although contributors appear to focus their attention on posts with higher GPT-estimated misleadingness, their fact-checking may simultaneously be biased toward content that is easier to assess. Such a focus can improve efficiency and support the production of high-quality notes, but it also risks overlooking politically salient or ambiguous posts that requestors perceive as urgent. Addressing this gap may require hybrid models of collaboration, where professional fact-checkers complement contributors by targeting content that is more ambiguous, emergent, or difficult to verify~\cite{augenstein2025community}.

\textbf{Partisan asymmetries between requests and community notes.}
Notably, the human-centered fact-checking approaches---whether expert-based or crowdsourced---cannot fully avoid political bias in the selection of targets~\cite{chuai2025political,pilarski2024community,shin2017partisan}. Our results reveal significant partisan asymmetries between reuqested posts and those that ultimately receive community notes. Previous work has shown that users preferentially challenge content authored by those with opposite partisan leanings, and Republicans are flagged more often than Democrats for sharing misinformation on \X's Community Notes~\cite{renault2025republicans,kuuse2025crowdsourced,allen2022birds}. Our findings echo this pattern: among requested posts, those authored by Republicans are more likely to receive community notes than those authored by Democrats. This asymmetry raises an important interpretive challenge. On the one hand, it may reflect partisan selection biases, with anonymized contributors and requestors each applying their own subjective judgments when deciding what to fact-check~\cite{martel2025political}. On the other hand, it could mirror underlying asymmetries in the misinformation landscape, where different political groups are disproportionately represented in the spread of misleading content. The coexistence of such tendencies highlights the complexity of maintaining balance and fairness in decentralized fact-checking systems like Community Notes program.

\textbf{Response of top writers to requests.}
Although the direct impact of the request feature on fostering note writing and coverage remains modest, requests appear to shift the attention of top writers: they are more likely to engage with political posts and focus on content with higher estimated misleadingness. By concentrating on posts with greater misinformation potential, request-fostered notes from top writers are evaluated as more helpful and less polarized than other notes on requested posts. In this sense, contributors play a crucial role as gatekeepers, directing community fact-checking toward high-risk content. 

At the same time, the request feature indirectly amplifies the influence of these contributors in shaping what gets fact-checked. A small elite of writers (22.2\%) is responsible for nearly half of all community notes (49.7\%), underscoring the uneven distribution of labor characteristic of volunteer-based systems. This concentration of fact-checking activity raises concerns about scalability and resilience: if the system relies disproportionately on a narrow core of expert-like participants, its long-term sustainability may be vulnerable to burnout, disengagement, or shifts in contributor incentives~\cite{augenstein2025community}. Moreover, while the reliance on elite contributors enhances quality and consensus, it also complicates the platform's vision of broad, community-driven participation. Rather than democratizing fact-checking, requests may consolidate authority in the hands of a few, blurring the line between peer production and expert review.

\textbf{Reliability and vulnerability of the note selection algorithm.}
We provide a comprehensive analysis of how the helpfulness and polarization of community notes are shaped by both content and source factors. Posts with higher estimated misleadingness and authored by accounts with greater misinformation exposure are more likely to elicit helpful annotations. This finding supports the view that Community Notes can function as a reliable mechanism for surfacing accurate fact-checks and correcting misinformation~\cite{allen2024characteristics}.

However, our findings on polarization also highlight a critical vulnerability. Notes on political or opinionated content tend to provoke divided evaluations, even when they cite high-quality domains, which resonates with previous research~\cite{toyoda2025understanding}. Such polarization does not necessarily imply inaccuracy, but can reflect partisan disagreement over credibility, legitimacy of sources, or interpretive framing~\cite{bouchaud2025algorithmic}. Therefore, the Community Notes system faces a significant challenge: factually accurate annotations may still fail to gain consensus if they are evaluated through polarized lenses. Addressing this limitation points to the need to improve the note selection algorithm to better balance factual accuracy with the practical requirement of fostering consensus among heterogeneous communities.

\subsection{Practical Implications}
Our findings provide several actionable insights for the design of community-based fact-checking systems on social media platforms, such as \X's Community Notes. 

\textbf{Reducing the delay of request alerts.} Requests currently function less as a mechanism to increase overall note volume and coverage, partly due to differences in selection patterns between contributors and requestors, and partly due to delays in reaching the request alert threshold relative to note generation. Although the first request may appear early in a post's diffusion, it often takes considerably longer to accumulate enough requests to trigger the alert. To reduce these delays, the request mechanism could be adapted to customize thresholds based on requestor reputation, granting lower thresholds to high-reputation users while maintaining safeguards for others. For example, \X is already testing a system that computes helpfulness scores for requestors, such that users with higher scores require fewer co-requests before their note requests are surfaced to contributors~\cite{cn2025request}.

\textbf{Leveraging LLMs to mitigate contributor bias and workload.}
Given recent advances in LLMs, these models could be incorporated into the Community Notes system to automatically write notes for requested posts with high estimated misleadingness. Such an integration would help mitigate the potential selection bias in note writers and also alleviate their workload. For instance, \X has opened the AI Note Writer API to developers, enabling the design of tools that assist in writing notes on the reuqested posts~\cite{cn2025ainote}. Beyond writing notes on posts requested by users, LLM-based tools can be expanded to provide proactive support across the platform---for example, issuing warnings to authors when they attempt to publish potentially misleading posts, reviewing posts after publication, and surfacing content with high misleadingness for community evaluation. Together, these applications would help broader and more uniform coverage of online content, complementing the request feature and strengthening the scability of community-based fact-checking.

\textbf{Addressing polarization and improving consensus.}
The current note selection algorithm faces a significant challenge: community notes, even when factually accurate, on posts related to politics and from high-influence users (\eg, verified accounts with many followers) tend to provoke polarized evaluations and fail to reach consensus among heterogeneous communities~\cite{bouchaud2025algorithmic}. To address this challenge, platforms could consider two directions for improvement. (i) Introducing an review layer by experts or contributors could help access highly polarized but potentially helpful notes, guiding decisions on which annotations to display~\cite{augenstein2025community}. In addition, the algorithm could assign higher weights to ratings from top contributors when aggregating evaluations. (ii) Our findings indicate that notes citing unbiased sources are evaluated as more helpful and less polarized, consistent with prvious research~\cite{solovev2025references}. Given this, platforms could provide contributor guidelines or incentives to encourage the use of neutral sources. Furthermore, LLMs can be leveraged to synthesize existing community notes, improve clarity, and then foster consensus-building across diverse user communities~\cite{de2025supernotes}.

Taken together, these actionable insights can inform the design and enhancement of community-based fact-checking systems. In particular, LLMs have promise for transforming the Community Notes system in streamlining request processing and assisting in note generation, thereby helping the system remain both reliable and responsive to user demands~\cite{li2025scaling}.

\subsection{Limitations and Future Work}
Our study has several limitations and offers directions for future research. First, our analysis is based on observational data from \X, which constrains causal inferences. Although we examine various characteristics---related to posts, their authors, and associated notes---and identify their associations with note outcomes, we cannot definitively determine whether certain factors directly influence contributors' decisions to write notes and causally lead to higher helpfulness or lower polarization in the resulting notes. For instance, in our post-level analysis (Fig.~\ref{fig:receive_note_coefs}), latent factors---such as unobserved contributor-level confounding and the inherent difficulty of fact-checking a post---are not directly measured or controlled for. These factors may influence both contributors' selection behavior and the quality of resulting notes. Future studies could explore causal relationships and clarify underlying mechanisms through controlled survey experiments or platform-supported interventions. Second, while our GPT-based approach provides scalable assessments of content types and potential misinformation, it remains subject to hallucinations and model-specific biases~\cite{augenstein2024factuality}, and may not fully capture emerging or highly complex misleading content~\cite{costabile2025assessing}. To mitigate these concerns, we conduct comprehensive validations using both assistant-annotated and expert-annotated datasets to demonstrate the reliability of GPT-generated outputs for our analysis. With continued advances in LLM-based fact-checking approaches, annotation accuracy is expected to further improve. Future work is encouraged to validate our findings when more advanced models are available. Third, given that the top-writer status is not available in our dataset, we cannot perfectly distinguish request-fostered notes from contributor-driven notes. Nevertheless, our estimation method ensures that the effect of the request feature in fostering note writing is not underestimated. Finally, contributors and requestors are anonymized in the Community Notes system, preventing the analysis of individual-level behavior, motivation, or expertise. While this anonymization can reduce rater bias, it also limits understanding of how personal experience or identity may influence content selection and note quality. Future work could explore these dynamics in contexts where such information can be collected ethically through survey experiments or accessed under appropriate privacy safeguards on platforms.

\section{Ethics Statement}
This research has received ethical approval from the Ethics Review Panel of the University of Luxembourg (ref. ERP 23-053 REMEDIS). 
All analyses are based on publicly available data. We declare no competing interests.

\begin{acks}
This research is supported by the Luxembourg National Research Fund (FNR) and Belgian National Fund for Scientific Research (FNRS), as part of the project REgulatory Solutions to MitigatE DISinformation, grant ref. INTER\_FNRS\_21\_16554939\_REMEDIS.
\end{acks}

\bibliographystyle{ACM-Reference-Format}
\bibliography{refs}

\newpage
\appendix

\begin{center}
    \huge \textbf{Supplementary Materials}\\
\end{center}

\renewcommand\thetable{S\arabic{table}}
\setcounter{table}{0}
\renewcommand\thefigure{S\arabic{figure}}
\setcounter{figure}{0}
\renewcommand\thesection{S\arabic{section}}
\setcounter{section}{0}

\section{Misleading Criteria}
\label{sec:misleading_criteria}
\begin{figure}[H]
    \centering
    \fbox{\includegraphics[width=.9\linewidth]{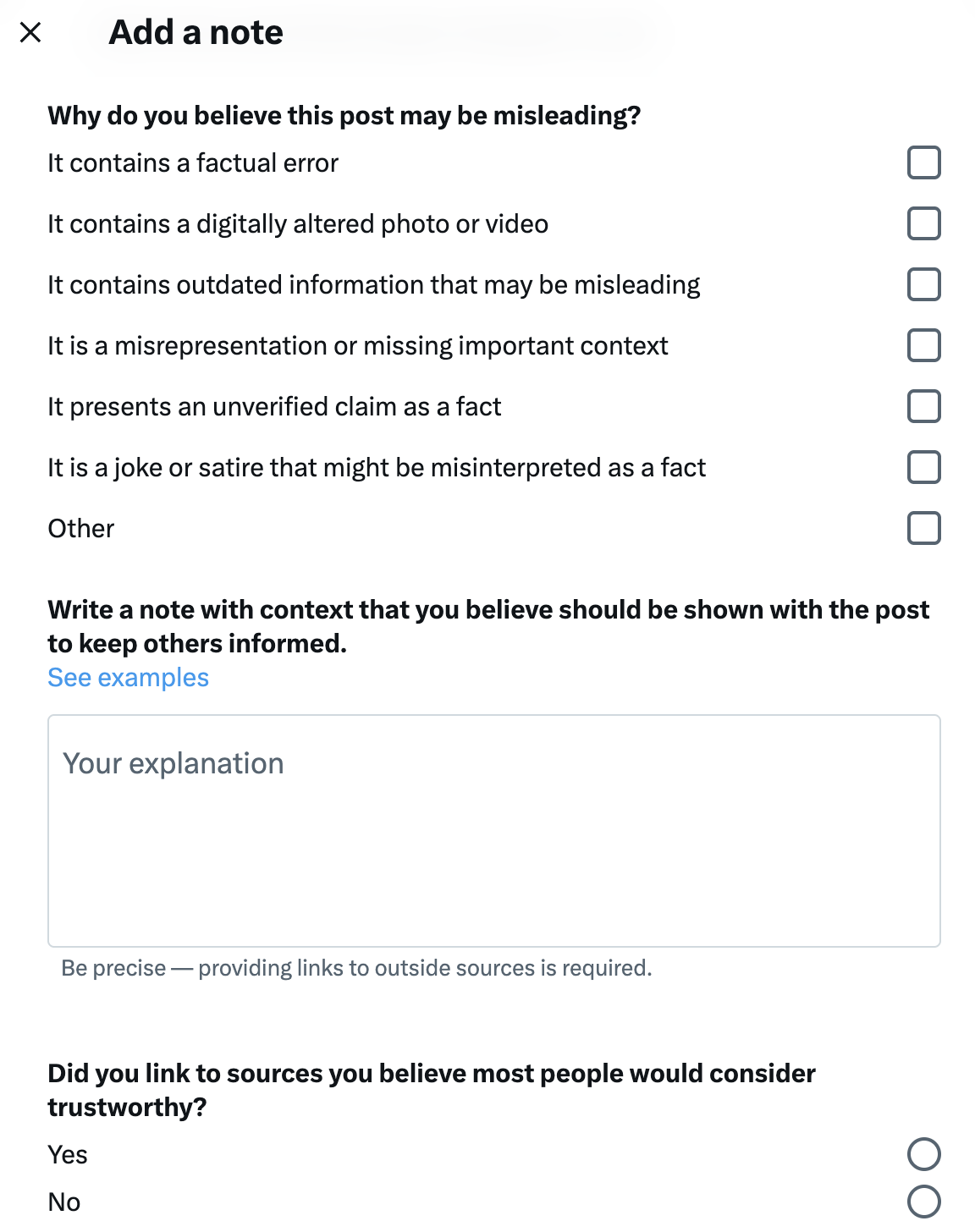}}
    \caption{The note-writing interface for Community Notes contributors, showing the specific criteria for labeling posts as misleading. Contributors are also required to provide explanations and cite external sources to justify their assessments.}
    \label{fig:note_page}
    \Description{}
\end{figure}

The criteria for labeling misleading posts in \X's Community Notes system is shown in Fig.~\ref{fig:note_page}. The listed criteria suggest that the Community Notes system aims to provide corrective or contextual information for not only factually incorrect content or alter images and videos, but also posts that present accurate information in a misleading manner, such as, through omission of context, exaggeration, or distortion of the original meaning. This is consistent with the definition of misinformation in our study. Therefore, we consider misleading posts annotated by community notes on \X to be manifestations of misinformation.

\section{GPT Prompt}
\label{sec:gpt_prompts}
The full GPT-4.1 prompt used in our analysis is provided below. In addition to the definition of misinformation, we adopt the definitions of claim and opinion in a previous study~\cite{ni2024afacta}.

\begin{quote}
You are an assistant that evaluates tweet content for two independent dimensions: content type and misleadingness.

\#\# Dimension 1: Content Type

Assess whether the tweet expresses a factual/verifiable claim and/or a subjective opinion.

\#\#\# Claim

A factual claim is a statement that explicitly presents some verifiable facts. 
Statements with subjective components like opinions can also be factual claims if they explicitly present objectively verifiable facts.

Indicators that a tweet contains a factual claim:\\
- Mentioning somebody (including the speaker) did or is doing something specific and objective\\
- Quoting quantities, statistics, and data\\
- Claiming a correlation or causation\\
- Assertion of existing laws or rules of operation\\
- Pledging a specific future plan or making specific predictions about future

\#\#\# Opinion

An opinion is a judgment based on facts, an attempt to draw a reasonable conclusion from factual evidence. 
While the underlying facts can be verified, the derived opinion remains subjective and is not universally verifiable.

\#\# Dimension 2: Misleadingness

Assess how misleading the tweet is, i.e., its potential to express misinformation, regardless of intent.

\#\#\#Definition

Misinformation refers to false or misleading information that is contradicted by empirical evidence or inconsistent with common or expert consensus, without imposing assumptions on the intent or the format of the content

Indicators of misinformation or misleadingness:\\
- It contains a factual error\\
- It contains a digitally altered photo or video\\
- It contains outdated information that may be misleading\\
- It is a misrepresentation or missing important context\\
- It presents an unverified claim as a fact\\
- It is a joke or staire that might be misinterpreted as a fact

\#\# Output Format

For each input tweet, return a single-line JSON string with the following structure:

\{\\
``type\_scores'': \{\\
``claim'': float (0 to 1),    \# Degree to which the tweet presents a factual/verifiable claim\\
``opinion'': float (0 to 1)   \# Degree to which the tweet expresses a subjective opinion\\
\},\\ 
``misleadingness'': float (0 to 1)  \# How misleading the tweet is, where 0 = not misleading, 1 = extremely misleading\\
\}

\#\# Guidelines:

- "claim" and "opinion" scores are independent and do not need to sum to 1.\\
- Only return the JSON string. Do not include any explanation or additional text.
\end{quote}

\section{GPT Validation Using PolitiFact}
\label{sec:gpt_politifact}

In addition to the manual validation conducted with student assistants, we further evaluate the performance of our GPT-based approach using expert-verified data from PolitiFact. Specifically, we collect all statements fact-checked by PolitiFact between July 15, 2024 and June 4, 2025, aligning with the time window of our dataset of requested posts. We obtain \num{1320} statements in total and retain those rated as (mostly) false ($=$ \num{1016}) or (mostly) true ($=$ \num{49}), resulting in \num{1065} statements for evaluation. Applying the same GPT-based annotation pipeline, we compare GPT assessments with PolitiFact's expert verdicts. GPT achieves a high performance ($\var{Accuracy}=$ 0.801; $\var{Weighted~F1}=$ 0.857), consistent with the results of our manual validation. Notably, despite the substantial class imbalance toward false statements, GPT attains comparable recall for true ($\var{Recall}=0.816$) and false statements ($\var{Recall}=0.800$), demonstrating its effectiveness in identifying both accurate and misleading content.

\section{Likelihood of Receiving Community Notes for Political Posts}
\label{sec:robustness_political}
\begin{figure*}
    \centering
    \includegraphics[width=\linewidth]{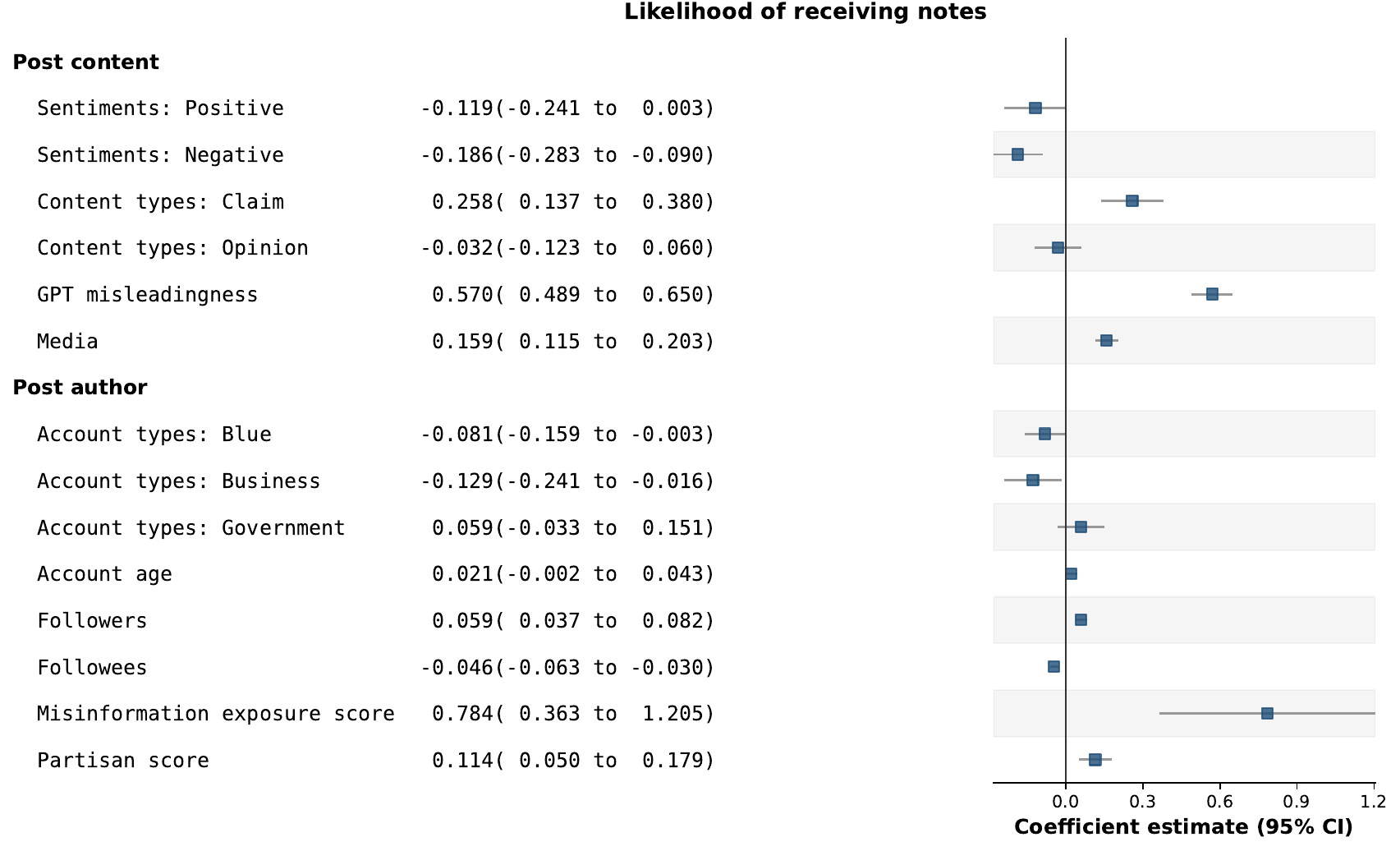}
    \caption{The estimation results for the logistic regression model predicting the likelihood that a requested political post receives a community note. Shown are coefficient estimates with 95\% CIs. The coefficients for misinformation exposure score and partisan score are estimated based on the subset of posts containing the corresponding information. The number of words in each post is controlled during estimation but omitted in visualization for better readability. Continuous independent variables---word count, account age, number of followers, and number of followees---are z-standardized before estimation to facilitate interpretation.}
    \label{fig:receive_note_coefs_political}
    \Description{}
\end{figure*}

Fig.~\ref{fig:receive_note_coefs_political} presents the estimation results for political post regarding the likelihood of receiving community notes. The results remain robust and consistent with our main findings.

\section{Robustness Check for Top Writers' Post Selection}
\label{sec:robustness_top_writers}
\begin{figure*}
    \centering
    \includegraphics[width=\linewidth]{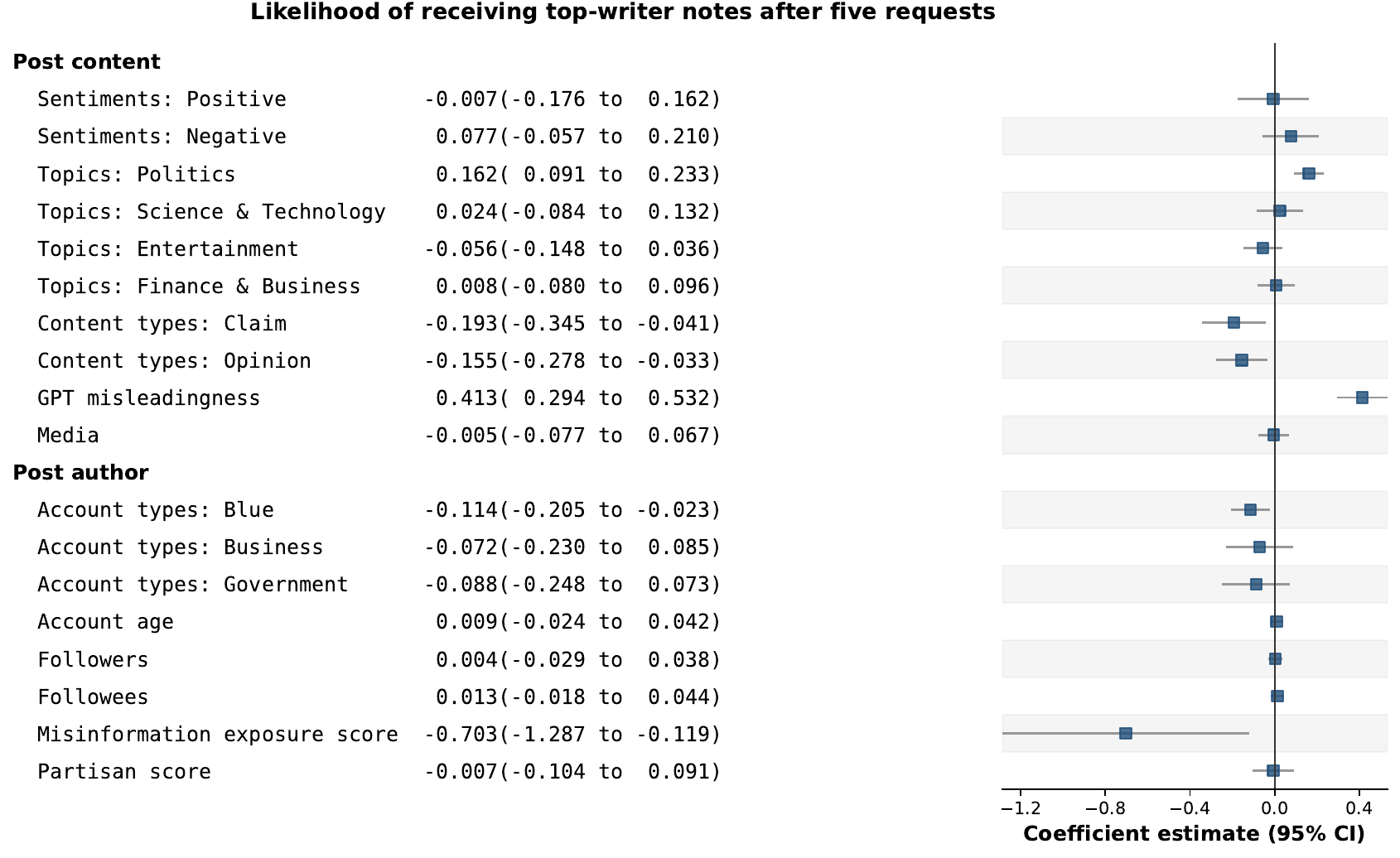}
    \caption{The estimation results for the logistic regression model predicting the likelihood that a requested post receives a top-writer note after the request threshold, \ie, request-fostered notes. Shown are coefficient estimates with 95\% CIs. The coefficients for misinformation exposure score and partisan score are estimated based on the subset of posts containing the corresponding information. The number of words in each post is controlled during estimation but omitted in visualization for better readability. Continuous independent variables---word count, account age, number of followers, and number of followees---are z-standardized before estimation to facilitate interpretation.}
    \label{fig:receive_note_coefs_top_writer}
    \Description{}
\end{figure*}

In the main paper, we find that top writers are more likely to select political posts and posts with higher potential misleadingness after the request threshold, compared to before the request threshold. To ensure the robustness of our findings and reduce potential unobserved confounding related to potential heterogeneity among top writers, we restrict the analysis to posts selected by the same set of top writers before and after the request threshold. The estimation results are presented in Fig.~\ref{fig:receive_note_coefs_top_writer}. The estimates for political posts ($\var{coef.}=0.162$, $\var{odds~ratio}=1.176$, $p<0.001$) and misleadingness ($\var{coef.}=0.413$, $\var{odds~ratio}=1.511$, $p<0.001$) remain consistent with our findings, further supporting that requests shift contributors' attention toward political content and posts with higher misleadingness.

\section{Robustness Check for Likelihood of Receiving Community Notes}
\label{sec:robustness_likelihood}

\begin{figure*}
    \centering
    \includegraphics[width=\linewidth]{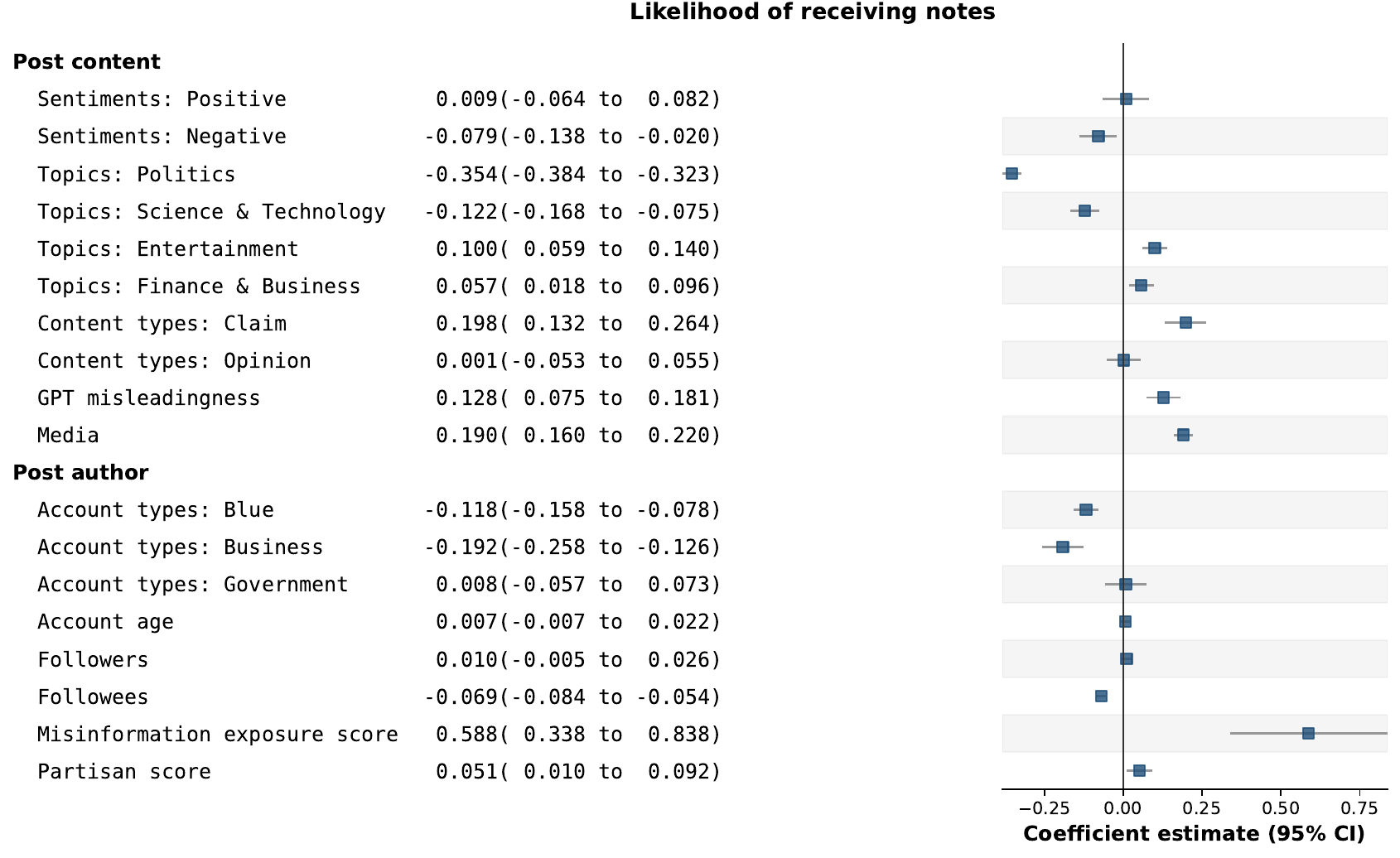}
    \caption{The estimation results for the logistic regression model predicting the likelihood that a requested post receives a community note. Shown are coefficient estimates with 95\% Confidence Intervals (CIs). Posts with request-fostered notes are omitted during estimation. The number of words in each post is controlled during estimation but omitted in visualization for better readability. Continuous independent variables---word count, account age, number of followers, and number of followees---are z-standardized before estimation to facilitate interpretation.}
    \label{fig:receive_note_coefs_robust}
    \Description{}
\end{figure*}

To further examine how Community Notes contributors differ from requestors in post selection and mitigate confounding from request-driven activity, we exclude posts with request-fostered notes identified in Section~\ref{sec:request_timing} and repeat our analysis on the likelihood of receiving community notes for the requested posts. The estimation results are reported in Fig.~\ref{fig:receive_note_coefs_robust}, and they remain robust and consistent with our main findings.

\end{document}